\documentclass[12pt]{article}
\usepackage{amsmath,amssymb,epsfig}
\usepackage{cancel}
\usepackage{color}

\makeatletter

\@addtoreset{equation}{section}
\makeatother

\begin{document}

\title{\vbox{
\baselineskip 14pt
\hfill \hbox{\normalsize KUNS-2497 }\\
\hfill \hbox{\normalsize EPHOU-14011}\\
\hfill \hbox{\normalsize YITP-14-40}
} \vskip 1.7cm
\bf Revisiting Discrete Dark Matter Model:\\
$\theta_{13}\neq 0$ and $\nu_R$ Dark Matter\vskip 0.5cm
}
\author{
Yuta~Hamada$^1$, \ \
Tatsuo~Kobayashi$^{1,2}$, \ \
Atsushi~Ogasahara$^1$, \ \  \\
Yuji~Omura$^3$, \ \
Fumihiro Takayama$^4$ 
 \ and \
Daiki~Yasuhara$^1$
\\*[20pt]
$^1${\it \normalsize 
Department of Physics, Kyoto University, 
Kyoto 606-8502, Japan }\\
$^2${\it \normalsize Department of Physics, Hokkaido University, Sapporo 060-0810, Japan}\\
$^3${\it \normalsize Department of Physics, Nagoya University, Nagoya 464-8602, Japan}\\
$^4${\it \normalsize Yukawa Institute for Theoretical Physics, Kyoto University, 
Kyoto 606-8502, Japan }
 \\*[50pt]}

\date{
\centerline{\small \bf Abstract}
\begin{minipage}{0.9\linewidth}
\medskip 
\medskip 
\small
We revisit the discrete dark matter model with the $A_4$ flavor symmetry originally introduced by M.Hirsch {\it et.al}. 
We show that radiative corrections can lead to non-zero $\theta_{13}$ and the non-zero mass for the lightest neutrino. 
We find an interesting relation among neutrino mixing parameters and it indicates the sizable deviation of $s_{23}$ from the maximal angle $s_{23}^2=1/2$ and the degenerate mass spectrum for neutrinos. 
Also we study the possibilities that the right-handed neutrino is a dark matter candidate.
Assuming that the thermal freeze-out explains observed dark matter abundance, TeV-scale right-handed neutrino and flavored scalar bosons are required. 
In such a case, the flavor symmetry plays an important role for the suppression of lepton flavor violating processes as well as for the stability of dark matter. 
We show that this scenario is viable within currently existing constraints from collider, low energy experiments and cosmological observations.
\end{minipage}
}

\newpage

\begin{titlepage}
\maketitle
\thispagestyle{empty}
\clearpage
%\tableofcontents
%\thispagestyle{empty}
\end{titlepage}

\renewcommand{\thefootnote}{\arabic{footnote}}
\setcounter{footnote}{0}

\section{Introduction}

The Higgs particle, which was the last missing piece of the Standard Model (SM), has been 
discovered, and other precision measurements have confirmed the SM.
However, still there are various mysteries on physics beyond the SM.
For example, the SM has many free parameters and most of them are relevant to 
the flavor sector, but we have not understood what the origin of complicated flavor structure is. 
On the other hand, astrophysical and cosmological observations tell the existence of dark matter, 
but we have not understood its origin in particle physics. 

The lepton sector has the specific form of mixing angles.
Two of them, $\theta_{12}$ and $\theta_{23}$, are large and the other, $\theta_{13}$, is of ${\cal O}(0.1)$.
In the limit, $\theta_{13} \rightarrow 0$, the Tri-bimaximal Ansatz  \cite{Harrison:2002er} 
was a good approximation for the lepton mixing matrix, i.e. the PMNS matrix.
The Tri-bimaximal matrix can be derived by using non-Abelian flavor 
symmetries such as $A_4$ and $S_4$ and assuming certain breaking patterns 
into Abelian symmetries, $Z_2$ and $Z_3$.
The exact Tri-bimaximal mixing is excluded by recent experiments, which 
showed $\theta_{13} \neq 0$ \cite{An:2012eh, Abe:2011fz, Ahn:2012nd, Abe:2011sj, Adamson:2011qu}.
However, the above approach through the use of non-Abelian discrete flavor symmetries 
is still interesting to realize the lepton mixing angles with $\theta_{13} \neq 0$ 
as well as the quark mixing angles.
(See for reviews of models with non-Abelian flavor symmetries 
\cite{Altarelli:2010gt,Ishimori:2010au,King:2013eh}.)

Dark matter may have heavy mass and couple with the SM particle.
A certain symmetry, e.g. the R-parity in supersymmetric standard models, is useful 
to make dark matter stable against decays into the SM particles.
Thus, the origin of dark matter may be related to the flavor structure, 
in particular the lepton flavor structure, 
and a single non-Abelian discrete symmetry may be concerned with 
both the realization of the lepton mixing angles and the stabilization of dark matter.

Recently, such a possibility was studied in the so-called discrete dark matter model to relate the lepton flavor structure and the origin of dark matter in Refs.\cite{Hirsch:2010ru,Boucenna:2011tj}.
\footnote{See also \cite{Kajiyama:2010sb,Meloni:2011cc,Kajiyama:2011fe,Boucenna:2012qb}.}
The discrete dark matter model has the $A_4$ flavor symmetry and 
the $A_4$ symmetry is assumed to break to the $Z_2$ symmetry and to lead to 
the lepton masses and mixing angles.
All of the SM particles have the $Z_2$ even charge, but some of right-handed neutrinos 
and the extra Higgs scalars coupled with only the neutrinos have the $Z_2$ odd charge.
Thus, the lightest particle with the $Z_2$ odd charge must be stable.
In \cite{Hirsch:2010ru,Boucenna:2011tj}, the extra Higgs scalar is assumed to be  
a dark matter candidate.
It was shown that the model leads to $\theta_{13} =0$ and the inverted hierarchy of 
neutrino masses with $m_3 =0$.
One may obtain $\theta_{13} \neq 0$ by extending the model.

In this paper, we revisit the discrete dark matter model.
We will show that radiative corrections can lead to 
$\theta_{13} = {\cal O}(0.1)$ and $m_3 \neq 0$ even without extending 
the original discrete dark matter model.
Both the inverted and normal hierarchies are possible.
We also study the possibilities that the right-handed neutrino is 
a dark matter candidate in this model.
\footnote{See, e.g. for works on right-handed neutrino dark matter\cite{Asaka:2005an}.}
In such a scenario, the typical mass scale of the model is 
as low as ${\cal O}(100-1000)$GeV.
In general, experimental constraints such as lepton flavor violation experiments and collider bounds 
have already set a limit on the right-handed neutrinos and the extra Higgs scalars with such a mass scale. 
However, in our scenario, the breaking scale of $A_4$ is quite low. 
That leads to a characteristic phenomenology and 
the flavor symmetry is also helpful to evade the strong experimental constraints.

This paper is organized as follows.
In section 2, we  review the discrete dark matter model.
In section 3, we study radiative corrections on neutrino masses.
In section 4, we study the scenario that the right-handed neutrino 
is lighter than the extra scalar and a dark matter candidate.
Several phenomenological aspects of our scenario are also studied.
Section 5 is devoted to conclusion and discussion.
In Appendix A, we show group theoretical aspects of $A_{4}$.
In Appendix B, we write explicitly the scalar potential, and study the mass spectrum.
In Appendix C, we show in detail the neutrino mass matrix.
In Appendix D, we discuss radiative corrections in the neutrino masses.

\section{Discrete dark matter model}
 
In this section, we briefly review the discrete dark matter model proposed in Refs.\cite{Hirsch:2010ru,Boucenna:2011tj} 
to give a dark matter candidate and an explanation for the flavor structure of the lepton sector simultaneously.

\subsection{Model}

In this model, the $A_{4}$ group, which is the symmetry group of the tetrahedron, is adopted as the lepton flavor symmetry group.
A brief description of the $A_{4}$ group is given in appendix A. 
$A_{4}$ has four irreducible representations, that is, three singlets$({\bf 1,1',1''})$ and one triplet$({\bf 3})$.
Ingredients of the discrete dark matter model are assigned to symmetry group representations according to the table below.

\begin{table}[h]
 \begin{center}
  \begin{tabular}{|c|c|c|c|c|c|c|c|c|c|c|}
   \hline
                 & $L_{e}$   & $L_{\mu}$  & $L_{\tau}$  & $e_{R}^c$   & $\mu_{R}^c$   & $\tau_{R}^c$  & $\nu_R=(\nu_R^{1},\nu_R^{2},\nu_R^{3})$ & $N_{4}$   & $h$       & $\eta=(\eta_{1},\eta_{2},\eta_{3})$ \\ \hline
  $\rm{SU(2)_L}$ & ${\bf 2}$ & ${\bf 2}$  & ${\bf 2}$   & ${\bf 1}$ & ${\bf 1}$   & ${\bf 1}$   & ${\bf 1}$                               & ${\bf 1}$ & ${\bf 2}$ & ${\bf 2}$                           \\ 
  $A_{4}$        & ${\bf 1}$ & ${\bf 1'}$ & ${\bf 1''}$ & ${\bf 1}$ & ${\bf 1''}$ & ${\bf 1'}$  & ${\bf 3}$                               & ${\bf 1}$ & ${\bf 1}$ & ${\bf 3}$                           \\ 
   \hline
  \end{tabular}
 \end{center}
\end{table}

$L_{\alpha}(\alpha=e,\mu,\tau)$ represent $\rm{SU(2)_L}$ doublets composed of a left-handed charged lepton and a left-handed neutrino. 
$e_{R},\mu_{R},\tau_{R}$ are right-handed charged leptons. $h$ is the Higgs boson. 
Adding to these SM particles, right-handed neutrinos $\nu_R^{i}(i=1,2,3)$, $N_4$ and $\rm{SU(2)_L}$ doublet scalars $\eta_{j}(j=1,2,3)$ are introduced. Each of $\nu_R^i$ and $\eta_j$ are put together into $A_{4}$ triplets. 

Each term in the Lagrangian must be constructed to be $A_{4}$ invariant. 
See Appendix A to check how to multiply non trivial $A_{4}$ representations together into the trivial singlet. 
The terms responsible for mass matrices of charged leptons and neutrinos are given by,

\begin{eqnarray}
\label{yukawa}
 {\cal L}_{\rm Yukawa}&=&y_{e}\overline{L}_{e}e_{R}h+y_{\mu}\overline{L}_{\mu}\mu_{R}h+y_{\tau}\overline{L}_{\tau}\tau_{R}h \nonumber \\
  &&+y_{\nu}^e\overline{L}_{e}(\nu_R\tilde{\eta})_{1}+y_{\nu}^{\mu}\overline{L}_{\mu}(\nu_R\tilde{\eta})_{1''}+y_{\nu}^{\tau}\overline{L}_{\tau}(\nu_R\tilde{\eta})_{1'}\\
  &&+Y_{4}\overline{L}_{e}N_{4}\tilde{h}+M_N\overline{\nu^c_R}\nu_R+M_{4}\overline{N^c_4}N_{4}+\mbox{h.c.} .\nonumber
\end{eqnarray}

The potential of scalar bosons is given in Appendix B. One comment has to be addressed here. 
In this paper, we introduce the following $A_4$ soft breaking bilinear term,
\begin{eqnarray}
-m_{h\eta_1}^2 \eta_1^{\dag}h+\mbox{h.c.} ,
\end{eqnarray}
which was not considered in the original paper~\cite{Hirsch:2010ru,Boucenna:2011tj}. 
We will explain the motivation in section 4. 
We assume $m_{\eta}^2>0$ and $m_{h\eta_1}^2/m_{\eta}^2\ll 1$ in most of discussions below. 
Under this assumption $m_{\eta}^2>0$ and the existence of the soft term Eq.(2.2), $\eta$ can acquire their non-zero vacuum expectation values(VEVs) when electroweak(EW) symmetry is violated, 
while light or massless scalar modes do not arise because the degrees of freedom of EW vacuum degeneracy of scalar bosons coincide with the degrees of freedom of longitudinal modes of massive electroweak gauge bosons.

\subsection{Neutrino mass matrices at tree level}

When scalar bosons of this model gets VEVs such that
\begin{eqnarray}
\label{vev}
\langle h^{0} \rangle=v_{h}\neq0,\quad \langle \eta_{1}^{0} \rangle=v_{\eta}\neq0,\quad \langle \eta_{2,3}^{0} \rangle =0,
\end{eqnarray}
the neutrino Dirac mass matrix is given by
\begin{eqnarray}
\label{dirac}
 m_{D}=\left(
 \begin{array}{cccc}
 y_{\nu}^{e}v_{\eta} & 0 & 0 & Y_{4}v_{h} \\
 y_{\nu}^{\mu}v_{\eta} & 0 & 0 & 0 \\
 y_{\nu}^{\tau}v_{\eta} & 0 & 0 & 0 \\
 \end{array} 
 \right)
  \equiv 
 \left(
 \begin{array}{cccc}
 x_{1} & 0 & 0 & y_{1} \\
 x_{2} & 0 & 0 &   0   \\
 x_{3} & 0 & 0 &   0   \\
 \end{array}
 \right),
\end{eqnarray}
from (\ref{yukawa}). Similarly, the Majorana mass matrix of right-handed neutrinos is
\begin{eqnarray}
m_{R}=\left(
 \begin{array}{cccc}
 M_{N} &   0   &   0   &   0   \\
    0  & M_{N} &   0   &   0   \\
    0  &   0   & M_{N} &   0   \\
    0  &   0   &   0   & M_{4}
 \end{array}
 \right).
\end{eqnarray}

Then we can get the Majorana mass matrix of left-handed neutrinos from these matrices with type-I seesaw mechanism,
\begin{eqnarray}
\label{majorana}
 m_{\nu}\equiv-m_{D}m_{R}^{-1}m_{D}^{T}=
 \left(
 \begin{array}{ccc}
  \frac{x_{1}^{2}}{M_{N}}+\frac{y_{1}^{2}}{M_{4}} & \frac{x_{1}x_{2}}{M_{N}} & \frac{x_{1}x_{3}}{M_{N}} \\[2mm]
             \frac{x_{1}x_{2}}{M_{N}}             & \frac{x_{2}^{2}}{M_{N}}  & \frac{x_{2}x_{3}}{M_{N}} \\[2mm]
             \frac{x_{1}x_{3}}{M_{N}}             & \frac{x_{2}x_{3}}{M_{N}} & \frac{x_{3}^{2}}{M_{N}}  
 \end{array}
 \right)\equiv
 \left(
\begin{array}{ccc}
 Y^{2} &  AB   & AC \\
   AB  & B^{2} & BC \\
   AC  &  BC   & C^{2}
\end{array}
\right).
\end{eqnarray}
Here, parameters which determine matrix elements are defined as
\begin{eqnarray}
A,B,C=\frac{x_{1,2,3}}{\sqrt{M_{N}}},\quad Y^{2}=\frac{x_{1}^{2}}{M_{N}}+\frac{y_{1}^{2}}{M_{4}}.
\end{eqnarray}

We can see now why the $A_{4}$ singlet $N_{4}$ is needed. 
If we did not have $N_{4}$, the rank of (\ref{majorana}) would be one because of (\ref{dirac}),
and we would get a degenerate spectrum of the left-handed neutrino masses which is excluded by experiments.

Note that Eq.(\ref{yukawa}) leads to the diagonal mass matrix for the charged lepton sector.
Thus, the PMNS matrix is determined only by the structure of the neutrino mass matrix.

At the tree level, the Majorana mass of the lightest left-handed neutrino is zero because the rank of (\ref{majorana}) is two. 
The eigenvector corresponding to this zero eigenvalue is $(0,-C,B)^{T}/\sqrt{B^{2}+C^{2}}$, which means $\sin{\theta_{13}}=0$, $m_3=0$ when it is assumed to be the third column of the PMNS matrix. 
This case realizes the Inverted Hierarchy(IH) mass pattern. 

\subsection{Dark matter candidate}

In this scenario, the $A_{4}$ flavor symmetry is broken by the vacuum alignment in Eq.(\ref{vev}).
The residual symmetry is $Z_{2}$ generated by
\begin{eqnarray}
\left(\begin{array}{ccc}
1 &  0 &  0 \\
0 & -1 &  0 \\
0 &  0 & -1 
\end{array}\right),
\end{eqnarray}
and the second and the third components of the $A_{4}$ triplets become odd under this residual $Z_{2}$ symmetry. 
That is, $\eta_{2}, \eta_{3}, \nu_R^{2}$, and $\nu_R^{3}$ belong to the $Z_2$ odd sector after the $A_{4}$ flavor symmetry is broken to $Z_{2}$ while all the other ingredients of this model have the $Z_2$ even parity.
Thus, the lightest particle in the $Z_2$ odd sector is stable and a good candidate for dark matter.

\section{Neutrino masses and mixing angles}

In this section, we investigate whether or not this model can explain both observed neutrino mass hierarchy and lepton generation mixing including non-zero $\theta_{13}$.

The lepton flavor mixing matrix takes the form as $V_{\mbox{PMNS}}=U_l^{\dag}U_{\nu}$ where $U_l$ and $U_{\nu}$ are unitary matrices to diagonalize the charged lepton and neutrino mass matrices. 
In this model, the charged lepton Yukawa couplings take diagonal form in the $A_4$ irreducible representation basis, we could safely take $U_l$ to unit matrix as a good approximation and the physical lepton generation mixings arise only from the neutrino mixing matrix $U_{\nu}$. 
In this paper, to explain non-zero $\theta_{13}$, we consider the extension modifying only neutrino mixing matrix $U_{\nu}$ 
and we do not consider the modification of the charged lepton mixing matrix $U_l$ 
because we would like to leave the $Z_3$ structure in charged lepton sector suppressing lepton flavor violating processes which is discussed in the next section.

\begin{figure}[htbp]
\centering\leavevmode
\includegraphics[scale=0.7]{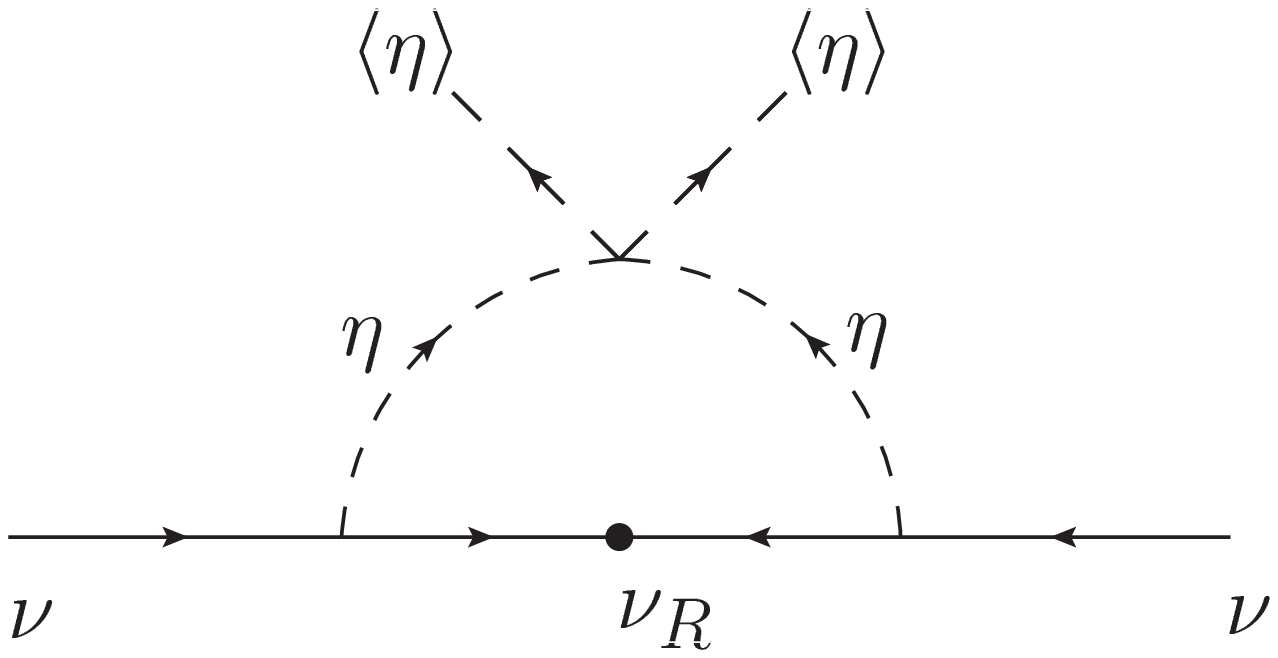}
\caption{The one-loop diagrams contributing to $A_4$ breaking neutrino masses under 
mass insertion approximations for $m_{h\eta_1}^2/m_{\eta}^2$.
\label{loop1}}
\end{figure}

\begin{figure}[htbp]
\centering\leavevmode
\includegraphics[scale=0.7]{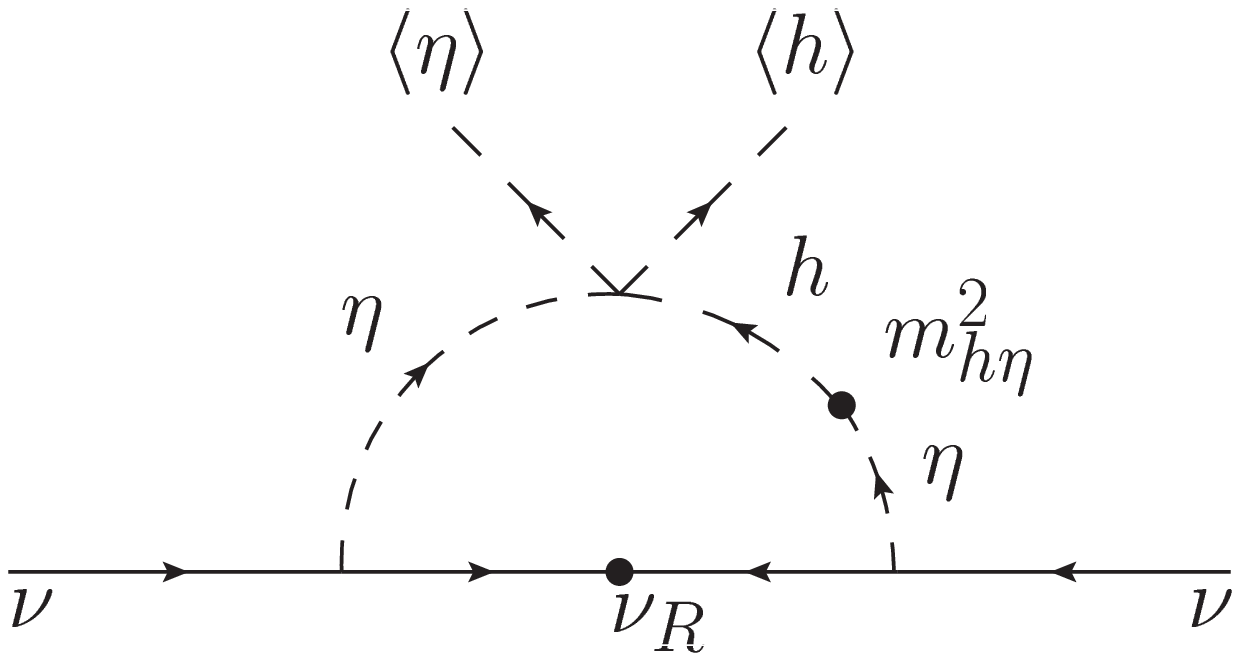}
\caption{The one-loop diagrams contributing to $A_4$ breaking neutrino masses under 
mass insertion approximations for $m_{h\eta_1}^2/m_{\eta}^2$.
\label{loop2a}}
\end{figure}

\begin{figure}[htbp]
\centering\leavevmode
\includegraphics[scale=0.7]{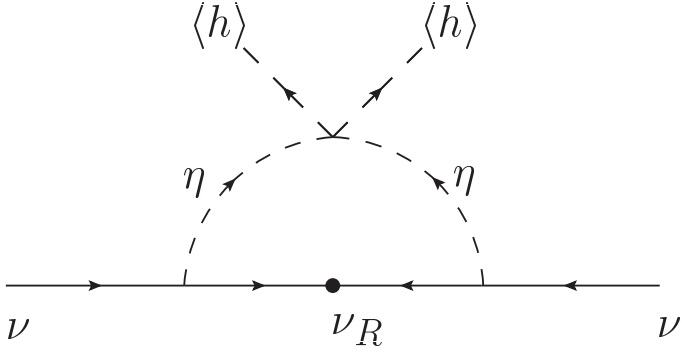}
\caption{The one-loop diagrams contributing to $A_4$ symmetric neutrino masses under 
mass insertion approximations for $m_{h\eta}^2/m_{\eta}^2$.
\label{loop3}}
\end{figure}

As we mentioned in the previous section, the tree-level contribution to neutrino mass with $N_4$ discussed in the original paper~\cite{Hirsch:2010ru,Boucenna:2011tj} can not achieve non-zero $\theta_{13}$. 
In this paper, we consider radiative corrections to neutrino masses which were not included in ~\cite{Hirsch:2010ru,Boucenna:2011tj}. 
The one-loop diagram contributing to neutrino masses are shown e.g in Fig.\ref{loop1}, Fig.\ref{loop2a} and Fig.\ref{loop3}. 
In general, the four point scalar boson interactions contain complex phases and can introduce CP phases to the neutrino mass matrix. 
See Appendix B for definitions of the quartic scalar couplings $\lambda_a$. 
Also we could add non-trivial singlet $N_5({\bf 1'})$ and $N_6({\bf 1''})$. The Yukawa interactions and the mass terms are as follows, 
\footnote{The modification for neutrino mass due to $N_5({\bf 1'})$ and $N_6({\bf 1''})$ was discussed at tree level in \cite{Meloni:2010sk}. } 
\begin{eqnarray}
&&L^{\mbox{Yukawa}}=Y_5\overline{L_{\mu}}N_{5}h+Y_6\overline{L_{\tau}}N_{6}h+\rm{h.c.},\\
&&L^{\mbox{mass}}=m_{N_5}\overline{N^c_5}N_6+\rm{h.c.}.
\end{eqnarray}


Since the rephasing of $N_{5,6}$ can not remove all phases of $Y_5$, $Y_6$ and $m_{N_5}$, these terms can be a source of CP phase in neutrino masses.  
The situation for $N_4$ is the same as the case of $N_{5,6}$.

This model generates neutrino masses in two different ways, seesaw mechanism and radiative corrections. 
See for detailed studies on radiative corrections Appendix D.
However the mass structures are classified into two types, $A_4$ symmetric $m^{\mbox{sym}}_{ij}$ and $A_4$ violating $m^{\mbox{break}}_{ij}$ parts in our current basis,
\begin{eqnarray}
(m_{\nu})_{ij}=m^{\mbox{sym}}_{ij}+m^{\mbox{break}}_{ij}.
\end{eqnarray} 
$A_4$ breaking parts are introduced by picking up $\langle \eta_{1}^{0} \rangle$ or $m_{h\eta_1}^2$. Taking the vacuum as 
$(\langle\eta_1\rangle,\langle\eta_2\rangle,\langle\eta_3\rangle)=(v_{\eta},0,0)$ to leave dark matter stable, the structure of the dominant part of $A_4$ breaking parts takes the following form~\footnote{ This form is a result of a condition imposed in our scalar potential. If the condition is relaxed, in general, it can be modified. See the detail in Appendix B and D.}
, 
\begin{eqnarray}
m^{\mbox{break}}_{ij}\simeq C_{\mbox{break}}y_{\nu}^i y_{\nu}^j \frac{v_{\eta}^2}{m_N},
\end{eqnarray}
with $C_{\mbox{break}}=1+C_{\mbox{rad}}^{\mbox{break}}$, where the first term arises from the tree-level seesaw contributions by right-handed neutrinos $\nu_R^i({\bf 3})$ exchanges 
and $C_{\mbox{rad}}^{\mbox{break}}=\rm{loop~factor}\times\lambda_{\Delta\eta=0}^{\mbox{break}}$ from corrections. 
Here, we named coupling constants which give contribution to radiative corrections$~\lambda_{\Delta\eta=0}^{\mbox{break}}$. 
The $A_4$ symmetric parts can be generated through the type-I seesaw mechanism by $N_i$ ($i=4,5,6$) exchanges or radiative corrections. 
$A_4$ symmetric nature reflects into the structure of mass matrix and the non-zero elements are,
\begin{eqnarray}
(m^{\mbox{sym}})_{11}&=& [C_{\mbox{rad}}^{\mbox{sym}}y_{\nu}^ey_{\nu}^e+Y_4Y_4\frac{M_N}{M_{N_4}}]\frac{v_h^2}{M_N},\\
(m^{\mbox{sym}})_{23}&=&(m^{\mbox{sym}})_{32}=[C_{\mbox{rad}}^{\mbox{sym}}y_{\nu}^{\mu}y_{\nu}^{\tau}+Y_5Y_6\frac{M_N}{M_{N_5}}]\frac{v_h^2}{M_N},
\end{eqnarray}
where $C_{\mbox{rad}}^{\mbox{sym}}=\rm{loop~factor}\times \lambda_{\Delta\eta=2}^{\mbox{sym}}$, and $\lambda_{\Delta\eta=2}^{\mbox{sym}}\sim \lambda_{11}$. 
$N_4({\bf1})$ seesaw contributes to the 11 entry of the $A_4$ symmetric parts and $N_5({\bf1'})$,$N_6({\bf1''})$ seesaw contributes to the 23 and 32 entries. 
The radiative corrections may contribute to all of 11 and 23, 32 entries. In general, $m^{\mbox{sym}}$, $m^{\mbox{break}}$ could be independent of each other. 
As one can see from (3.6), the contribution to neutrino mass matrix of $N_{5,6}$ 
and the $A_4$ symmetric parts of radiative correction enter the same  mass matrix elements. 
Then, if scalar potential is CP invariant, we see that the same form of the neutrino mass matix is obtained in both the original discrete dark matter model including radiative corrections without $N_{5,6}$ and the model with $N_{5,6}$ neglecting radiative corrections. On the other hand, if scalar potential contains CP phases, in general, radiative corrections can introduce more freedom than the case that $N_{5,6}$ are added and only tree level contributions are considered.~\footnote{For example, the tree level contributions due to $N_{5,6}$ can not change the form of $m^{\mbox{break}}$ given in Eq.(3.4) but radiative corrections in the general case of CP violating scalar potential may modify the form. See the detail in Appendix D.}

As we explain the detail in Appendix D, for the case that the scalar potentail has the invariance for ($\eta_2$,$\eta_3$) odd permutation which may be naturally realized e.g in the case of CP invariant scalar potential, we find the following general form of neutrino mass matrix in this model,~\footnote{This form of the neutrino mass matrix is identical to the one considered in \cite{Ahn:2013mva}}
\begin{eqnarray}
m_{\nu}=
\left( \begin{array}{ccc}
a^2+X_A & ab & ac\\
ab & b^2 & bc+X_B\\ 
ac & bc+X_B & c^2
\end{array} \right).
\end{eqnarray}
We will further investigate the phenomenological consequences below. We have five complex free parameters in the neutrino mass matrix. 
On the other hand, taking phase redefinition of $L_i$ ($i=e,\mu,\tau$), for example, we can remove the phases of $a$, $b$, $c$ and they can be taken as real numbers. 
Thus we have three real ($a$, $b$, $c$) and two complex ($X_A$, $X_B$) physical parameters. 
Then in such a basis, $X_A$ and $X_B$ can be regarded as two sources of CP phases which can not be removed by the field phase redefinition of $L_i$. 
If the all elements of $(m_{\nu})$ are real, the phase redefinition arguments in this model require that $(m_{\nu})_{22}/(m_{\nu})_{33} $, $((m_{\nu})_{11}-X_A)/(m_{\nu})_{22}$ are real positive numbers.

Notice that this model predicts one relation among the elements of the neutrino mass matrices,
\begin{eqnarray}
\frac{(m_{\nu})_{12}^2}{(m_{\nu})_{22}}=\frac{(m_{\nu})_{13}^2}{(m_{\nu})_{33}}.
\end{eqnarray}
In general, this condition is imposed on complex numbers of matrix elements. Then we have two conditions on real numbers of parameters, that is,
\begin{eqnarray}
\mbox{Re}\left(\frac{(m_{\nu})_{12}^2}{(m_{\nu})_{22}}-\frac{(m_{\nu})_{13}^2}{(m_{\nu})_{33}}\right)=\mbox{Im}\left(\frac{(m_{\nu})_{12}^2}{(m_{\nu})_{22}}-\frac{(m_{\nu})_{13}^2}{(m_{\nu})_{33}}\right)=0.
\end{eqnarray}
Notice that for any phase basis of $L_i$, the above conditions for real and imaginary parts have to be satisfied. 

The first question to be answered is whether this condition (3.9) is allowed or not in the current observational results. 
It restricts neutrino masses and mixing parameters, that is, we expect a relation among them as we will discuss it later. 
Taking neutrino masses as $|m_i|$ ($i=1,2,3$) and using the conventional form of the PMNS mixing matrix,
\begin{eqnarray}
&&U_{\mbox{PMNS}}=VP_{\nu},\\
&&~~~V=
\left(\begin{array}{ccc}
c_{12}c_{13} & s_{12}c_{13} & s_{13}e^{-i\delta}\\
-s_{12}c_{23}-c_{12}s_{13}s_{23}e^{i\delta} & c_{12}c_{23}-s_{12}s_{13}s_{23}e^{i\delta} & c_{13}s_{23}\\
s_{12}s_{23}-c_{12}s_{13}c_{23}e^{i\delta} & -c_{12}s_{23}-s_{12}s_{13}c_{23}e^{i\delta} & c_{13}c_{23}\\
\end{array}
\right),\\
&&~~~P_{\nu}=\left(\begin{array}{ccc}
1&0&0\\
0&e^{i\phi_2/2}&0\\
0&0&e^{i\phi_3/2}
\end{array}
\right),
\end{eqnarray}
where $s_{ij}=\sin{\theta_{ij}}$ and $c_{ij}=\cos{\theta_{ij}}$, we could relate the neutrino mass matrix to observed mixing parameters, that is, 
\begin{eqnarray}
(m_{\nu})=U_{\mbox{PMNS}}
\left(\begin{array}{ccc}
|m_1|&0&0\\
0&|m_2|&0\\
0&0&|m_3|
\end{array}
\right)
U^{T}_{\mbox{PMNS}}.
\end{eqnarray}
We list the concrete expressions for the neutrino mass matrix in Appendix C.
In Fig.\ref{Fig0}, we show the values of the observationally preferred mass matrix elements for the case of IH mass pattern as an example by varying observable values within 3 $\sigma$ of Table 1.
We see that there is a region where the above relation Eq.(3.8) is satisfied. 
In following discussions, regarding $m_2$ and $m_3$ as complex numbers, $m_2=|m_2|e^{i\phi_2}$ and $m_3=|m_3|e^{i\phi_3}$, we take $P_{\nu}=1$ without loss of generality.
 
Notice that the relation Eq.(3.8) has to be satisfied even in the case of previous studies~\cite{Hirsch:2010ru,Boucenna:2011tj} where $\theta _{13}=0$ is taken. 
We easily find that $s_{23}^2=1/2$, $s_{13}=0$, $e^{-i\delta}=e^{i\phi_1}=e^{i\phi_2}=1$ satisfy the relation Eq.(3.8) and it can realize the Tri-bimaximal mass pattern 
previously discussed in the original paper~\cite{Hirsch:2010ru,Boucenna:2011tj}. 
In the case of non-zero $\theta_{13}$, Eq.(3.8) requires $\delta m_{12}=m_2-m_1=0$ at $s_{23}^2=1/2$ according to the discussion in Appendix C. This is a trivial solution of Eq.(3.8).
We find the general solutions of Eq.(3.8) for non zero $\theta_{13}$ by shifting $\delta s_{23}$ and $\delta m_{12}$ from the trivial solution 
and the solution sensitively constrains deviation from $s_{23}^2=1/2$, $\delta s_{23}=s_{23}-\mbox{sgn}(s_{23})/\sqrt{2}$ as a function of other mixing parameters. 
This is an interesting prediction of this model. 
We give the exact form of $\delta s_{23}$ as a function of other mixing parameters in Eq. (C.18) of Appendix C. 
\begin{figure}[htbp]
\centering\leavevmode
\includegraphics[scale=1.5]{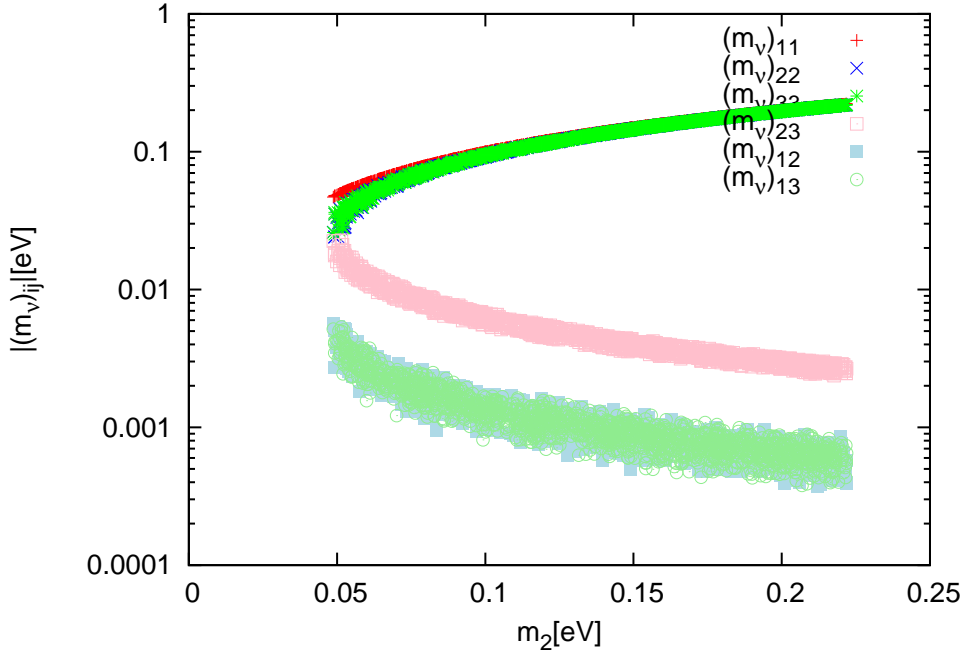}
\caption{$m_2$ and  observationally preferred $(m_{\nu})_{ij}$ in the case of 
IH mass pattern with $\sum m_{\nu}< 0.66\rm{eV}$~\cite{Ade:2013zuv}.
\label{Fig0}}
\end{figure}
Notice that $e^{i\delta}=e^{i\phi_1}=e^{i\phi_2}=1$ automatically satisfy the condition for the imaginary part of Eq.(3.9). 
First, we investigate the model implication to the neutrino mixing parameters under this phase condition for simplicity. 
Later we will relax this condition for phases.

\begin{table}[th]
\begin{center}
\begin{tabular}{|c|c|c|}
\hline
 Parameter                                 & $3 \sigma$ range       & best fit value \\
\hline
$\Delta m^2_{21}$ $( 10^{-5} {\rm eV}^2 )$ & $6.99 - 8.18$          & 7.54           \\
$|\Delta m^2|$ $( 10^{-3} {\rm eV}^2 )$    & $2.19-2.62(2.17-2.61)$ & 2.43(2.42)     \\
\hline
 $\sin^2 \theta_{12}$  & $0.259-0.359$                   & 0.307          \\
 $\sin^2 \theta_{23}$  & $0.331-0.637 (0.335-0.663)$     & 0.386(0.392)   \\
 $\sin^2 \theta_{13}$  & $0.0169-0.0313 (0.0171-0.0315)$ & 0.0241(0.0244) \\
\hline
\end{tabular}
\caption{
\label{table1}
{The $3 \sigma$ allowed ranges [19]. The values are in the case with $m_1<m_2<m_3$.
The values in bracket correspond to $m_3<m_1<m_2$. $\Delta m^2=m_3^2-(m_1^2+m_2^2)/2$ is defined.
}
}
\end{center}
\end{table}

\begin{figure}[htbp]
\centering\leavevmode
\includegraphics[scale=1.5]{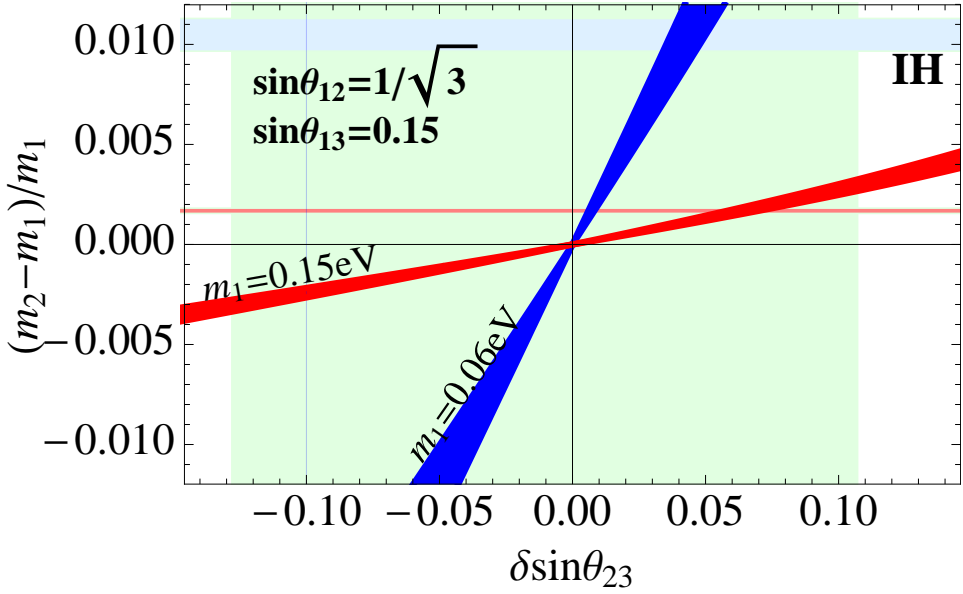}
\caption{$\delta s_{23}$ and $\delta m_{12}/m_1$ in the IH case with  $s_{23}>0$, $s_{12}=1/\sqrt{3}$ and $s_{13}=0.15$. 
The red (blue) line is the prediction of our $A_4$ model with $m_1=0.15(0.06)$eV, within $3 \sigma$ of $|\Delta m^2|$. 
The light green region for the $3\sigma$ allowed range of $\sin^2 \theta_{23}$, and the light pink (blue) band for the one of $\Delta m^2_{12}$ with $m_1=0.15(0.06)$eV respectively. \label{Fig1}}
\end{figure}

As for the case of IH mass pattern, observations require $\delta m_{12}^2/\delta m_{13}^2\sim 3\times 10^{-2}$ where $\delta m^2_{ij}=m_j^2-m_i^2$, and the mass difference $\delta m_{12}=m_2-m_1$ is always very small compared with $m_1$ and $m_2$ in this case. Near the observed values of mixing parameters, we approximately translate the relation Eq.(3.8) into the following form, 
\begin{eqnarray}
\delta m_{12}\simeq -\gamma\times 2\sqrt{2}\frac{s_{13}}{s_{12}c_{12}}\delta s_{23}\delta m_{13},
\end{eqnarray}
where $\gamma=m_1/(\delta m_{13}+2m_1)$ and $\delta m_{13}=m_3-m_1$. 
It is easy to see that this relation can be satisfied within the current observational results at $3\sigma$ level
\footnote{We used a global fit result~\cite{Globalfit2}. There are the other similar studies~\cite{Globalfit1}. 
These are consistent each others at 3$\sigma$ level, but there is a difference in the allowed regions within $2\sigma$ level due to the different treatment of observational data. 
There are recent developments measuring $s_{23}$ precisely. For examples, if we use T2K~\cite{Abe:2014ugx} seriously, then $s_{23}^2=1/2$ is still allowed enough. 
On the other hand, MINOS results~\cite{Adamson:2014vgd} seems a little bit disfavoring $s_{23}^3=1/2$.} 
and we find a tight correlation between the smallness of $\delta m_{12}$ and $\delta s_{23}$. 
By using the best fit values for masses and mixing parameters shown in Table~\ref{table1} and leaving $s_{23}$ as a free parameter, 
we could see that the maximal angle $s_{23}^2=1/2$ is excluded for non zero $\theta_{13}$ but $s_{23}^2$ still has to be close to $1/2$ 
and we find $\delta s_{23}\sim +0.015$ for the case of $\delta m_{13}\sim m_1$ ($m_1\sim 0.05\mbox{eV}$) and $\delta s_{23}\sim +0.06$ for the case  of $m_1>\delta m_{13}$ ($m_1>0.1\mbox{eV}$). 
By using the exact form of $\delta s_{23}$ Eq.(C.18) and varying the values of $s_{12}$, $s_{13}$ within current 3$\sigma$ errors of Table~\ref{table1}, we can still see the qualitatively same results as shown in Fig.~\ref{Fig1}.

\begin{figure}[htbp]
\centering\leavevmode
\includegraphics[scale=1.5]{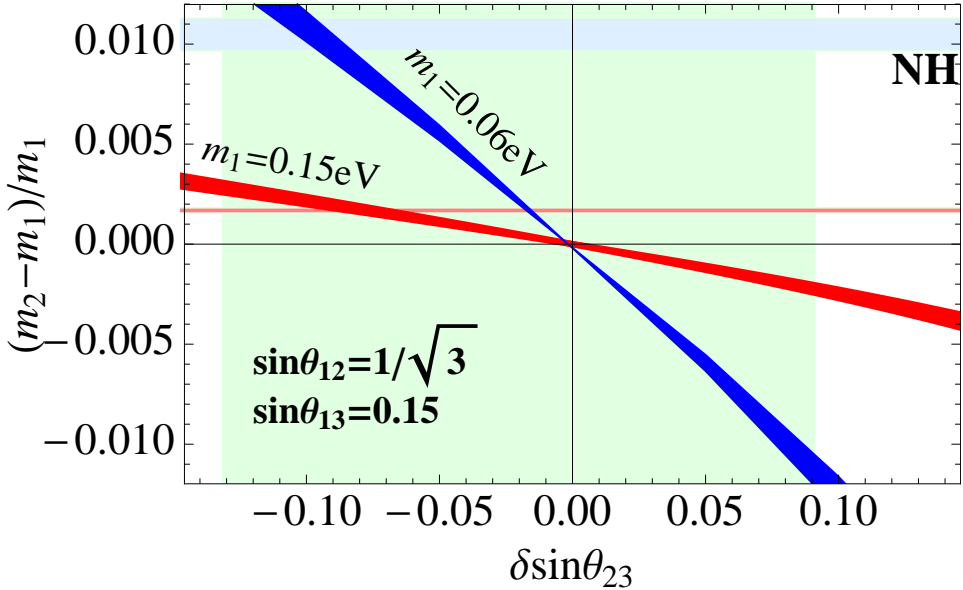}
\caption{$\delta s_{23}$ and $\delta m_{12}/m_1$ in the NH case with  $s_{23}>0$, $s_{12}=1/\sqrt{3}$ and $s_{13}=0.15$. 
The red (blue) line is the prediction of our $A_4$ model with $m_1=0.15(0.06)$eV, within $3 \sigma$ of $|\Delta m^2|$. 
The light green region for the $3\sigma$ allowed range of $\sin^2 \theta_{23}$, 
and the light pink (blue) band for the $\Delta m^2_{12}$ with $m_1=0.15(0.06)$eV respectively. \label{Fig2}}
\end{figure}

In a similar way,  we investigate the case of normal hierarchy (NH) mass pattern. 
In the case of $m_1>\delta m_{13}$ which realizes degenerate spectrum for three neutrinos, the mass hierarchy $\delta m_{12}\sim 3\times 10^{-2}\delta m_{13}$ is required by experimental results. 
In this case, we find the same approximated relation given in the previous IH case, Eq.(3.14).
The difference between the NH case and the IH cases is only the sign of $\delta m_{13}$.
The observed mass hierarchy and mixing angles require $\delta s_{23}\sim O(0.1)$ and we find that $\delta s_{23}\sim -0.06$ is preferred if we assume $m_1\gg\delta m_{13}$.
Notice that the sign of $\delta s_{23}$ is opposite to the IH case and the negative sign is preferred by the global fit of experimental data\cite{Globalfit2}.
Increasing the value of $\delta m_{13}/m_1$ up to $\sim 1$ , $\delta s_{23}$ increases up to $\sim 0.15$ and it reaches outside of the 3$\sigma$ allowed region of $\delta s_{23}>0.12$.
For $\delta m_{13}> m_1$, the approximation of Eq.(3.14) is not always valid and we numerically checked that for $m_1<0.04$eV, $\delta s_{23}$ reaches outside of allowed region of experimental data in the case.
This is again numerically confirmed in Fig.\ref{Fig2}.

\begin{figure}[htbp]
\centering\leavevmode
\includegraphics[scale=1.2]{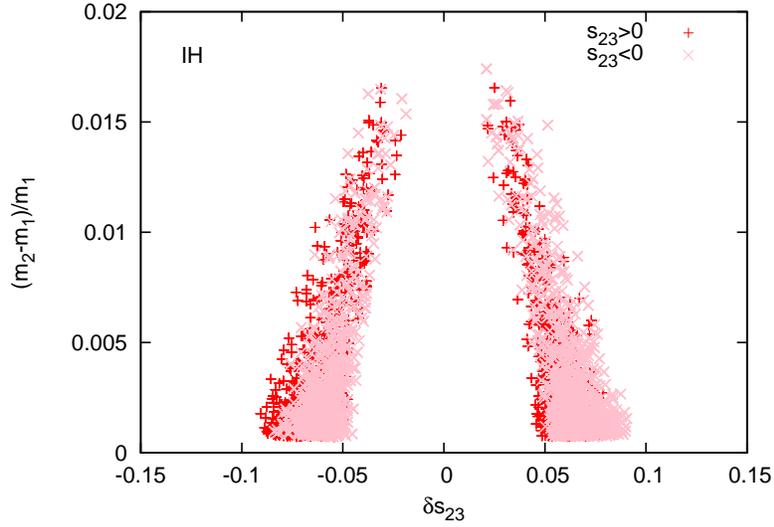}
\caption{The scatter plot showing the $A_4$-model-inspired allowed region for $\delta m_{12}$ and $\delta s_{23}$ in IH mass pattern. 
The mixing parameters $s_{12}$, $s_{13}$ are taken within the $3\sigma$ allowed range in Table 1.
Increasing $m_1$ obtains decreasing observationally preferred $\delta m_{12}/m_1$. \label{Fig3IH}}
\end{figure}

\begin{figure}[htbp]
\centering\leavevmode
\includegraphics[scale=1.2]{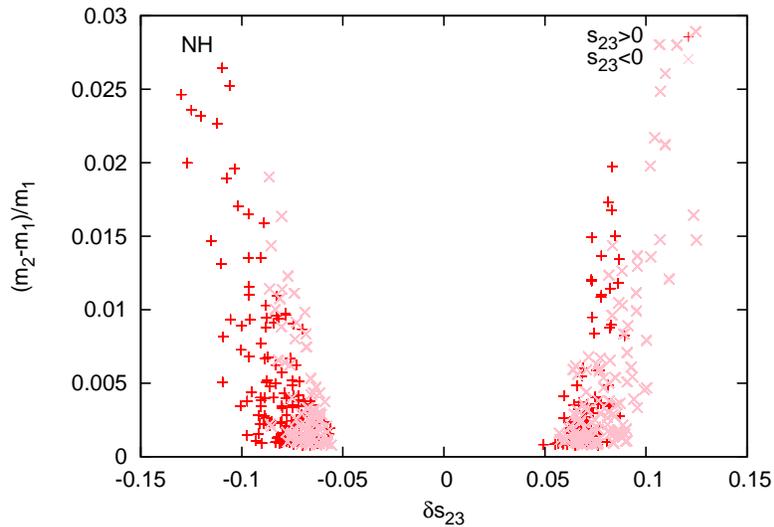}
\caption{The scatter plot showing the $A_4$-model-inspired allowed region for $\delta m_{12}$ and $\delta s_{23}$ in NH mass pattern. 
The mixing parameters $s_{12}$, $s_{13}$ are taken within the $3\sigma$ allowed range. 
Decreasing $m_1$ indicates increasing observationally preferred $\delta m_{12}/m_1$. \label{Fig3NH}}
\end{figure}

To see the above statements, we show the scatter plots for both IH (Fig.~\ref{Fig3IH}) and NH (Fig.~\ref{Fig3NH}) cases where all mixing parameters except for $s_{23}$ are varied within the 3$\sigma$ range given in Table 1.
For both NH and IH, non zero $\theta_{13}$ excludes the possibility of $s_{23}^2=1/2$, and the tight correlation of the smallness of $\delta s_{23}$ and $\delta m_{12}$ exists. This is a robust prediction of this model.
The deviation from $s_{23}$ obtain $0.01<|\delta s_{23}|\lesssim 0.1$ for $|\delta m_{13}|\lesssim m_1$ and increasing $\delta m_{13}/m_1$, $|\delta s_{23}|$ increases 
and it reaches outside of experimentally allowed range when we take $m_1\lesssim 0.03\mbox{eV}$ . 

Until now, we considered only the case that all Majorana phases and Dirac CP phase are trivial. 
Taking account for the effect of Majorana phases, for example, we can change the sign of $m_2$ and $m_3$, that is, taking $\phi_1=0, \pi$, $\phi_2=0, \pi$. 
In this case, the approximated form Eq.(3.14) is not always valid, especially for $m_2<0$ cases. 
We use Eq.(C.18) to determine $s_{23}$ satisfying condition Eq.(3.8) without any approximation, and estimate $\delta s_{23}$ for several combinations of the sign of $m_2$, $m_3$. 
We show the results in Fig.~\ref{ds23majorana}. Also, in Appendix C, we notice $\delta s_{23}\propto s_{12}s_{13}$. 
As the result, for the change of the sign of $s_{12}$, $s_{13}$, the flip of the sign of $s_{12}s_{13}$ causes the flip of the sign for $\delta s_{23}$. 
If we include Dirac CP phase $\delta$ for real $m_1$, $m_2$, $m_3$, $\sin\delta=0$ is one of the solution, which obtain $e^{-i\delta}=\pm 1$. 
The effect is identical to the effect of the sign flip of $s_{13}$.


\begin{figure}[htbp]
\centering\leavevmode
\includegraphics[scale=0.9]{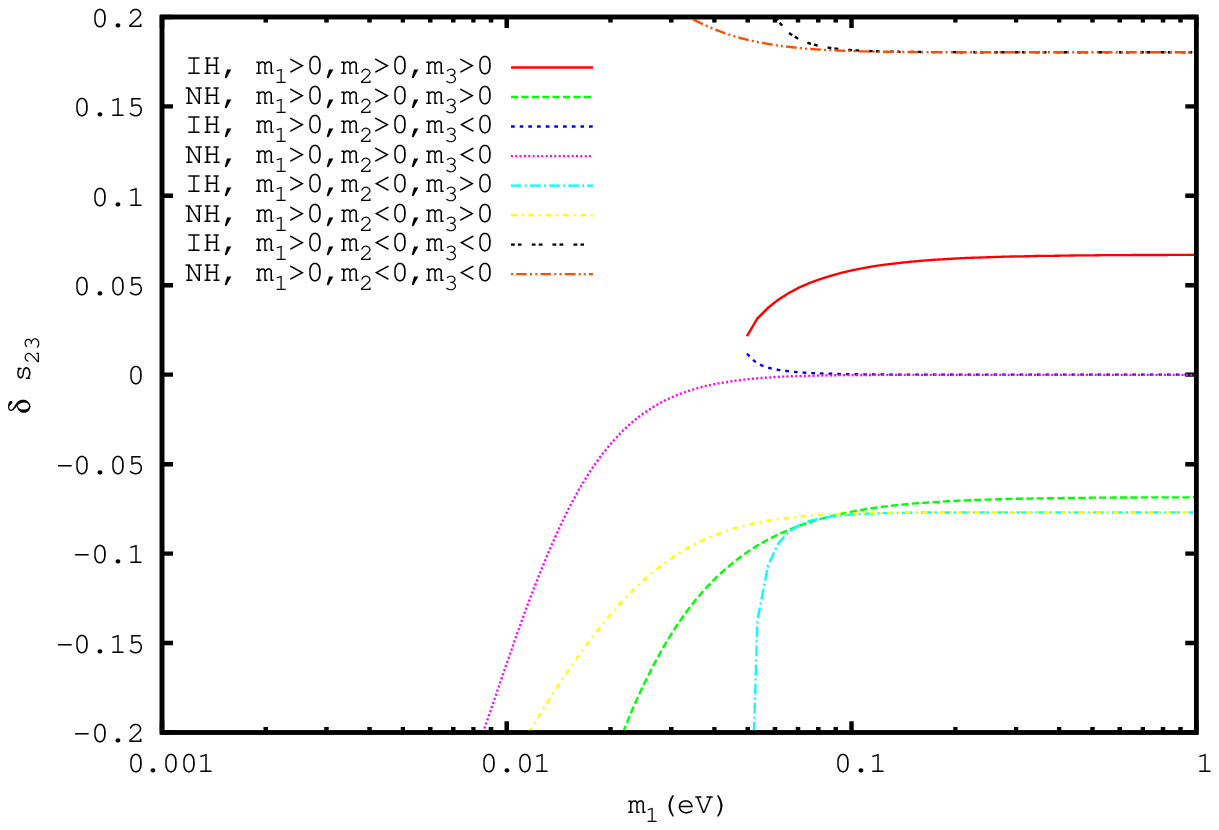}
\caption{We show the plot for $\delta s_{23}$ vs $m_1$ for the various cases when we take different Majorana phase (0 or $\pi$) for $m_2$ and $m_3$. 
We assumed $s_{12}>0$ and $s_{13}>0$ and we fixed $\delta m_{12}^2$, $\delta m_{13}^2$, $s_{12}^2$ and $s_{13}^2$ to the best fit values in Table 1.
When we flip the sign of $s_{13}$, $s_{12}$, the sign of $\delta s_{23}$ flips as sign of $s_{12}s_{13}$. 
As for the change of sign of $s_{23}$, the sign of $\delta s_{23}$ is unchanged. \label{ds23majorana}}
\end{figure}

\begin{figure}[htbp]
\centering\leavevmode
\includegraphics[scale=0.9]{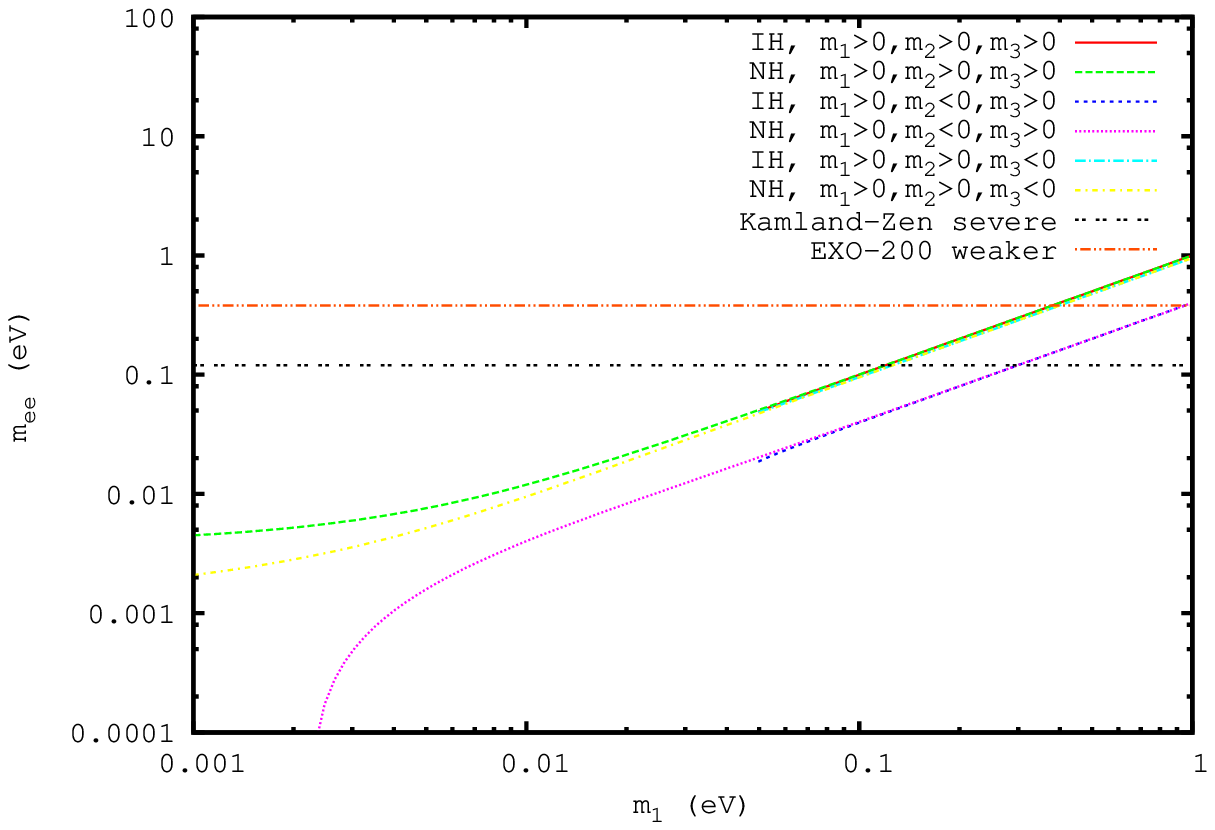}
\caption{We show the constraints on $m_1$ imposed by KamLAND-Zen and EXO-200 results for neutrinoless double beta decay. \label{doublebeta}}
\end{figure}

\begin{figure}[htbp]
\centering\leavevmode
\includegraphics[scale=0.9]{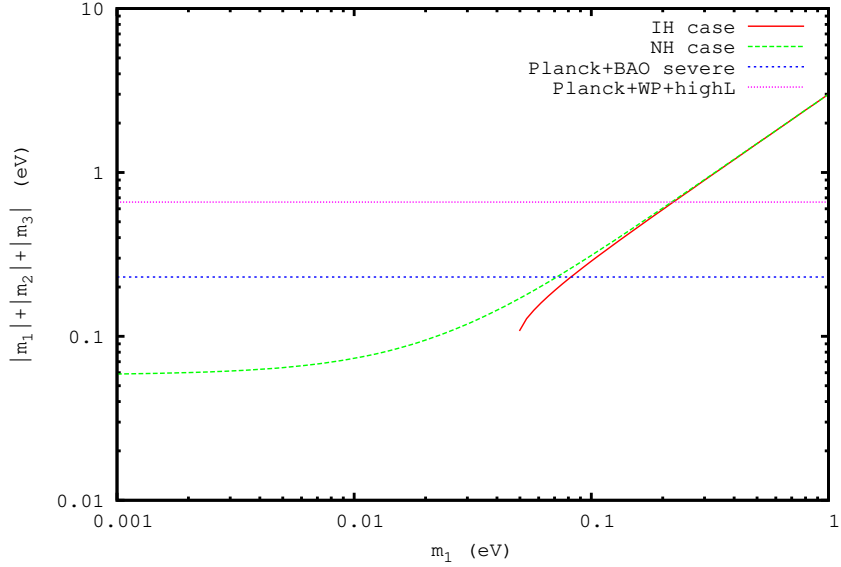}
\caption{We show the constraints on $m_1$ imposed by cosmological observation of Planck satellite. 
The lower horizontal line corresponds to the combined constraint with BAO and the upper horizontal line corresponds to the combined constraint with WMAP and high red shift survey. \label{cosmo}}
\end{figure}

From Fig.\ref{doublebeta}, we find that the solutions for IH and NH cases are allowed by the current neutrinoless double beta decay experiments~\cite{doublebeta}, 
and we may expect the observation or the exclusion of the large parts in future. 
In this degenerate mass spectrum, as we see in Fig.\ref{cosmo}, $m_1\gtrsim 0.07\mbox{eV} (\mbox{NH}), 0.08\mbox{eV}(\mbox{IH})$ faces a milder tension with the results of recent Planck CMB observation 
by seriously taking the BAO data, but it may be still allowed in general if we do not combine the Planck data with the BAO data~\cite{Ade:2013zuv}. 
Also variations of $N_{\nu \mbox{eff}}$ from the SM value may obtain milder constraints on $\sum m_{\nu}$~\cite{Ade:2013zuv}. 
However, too large $m_1>1\mbox{eV}$ has been already excluded by both the neutrinoless double beta decay experiments and the cosmological observations.


\begin{figure}[htbp]
\centering\leavevmode
\includegraphics[scale=0.9]{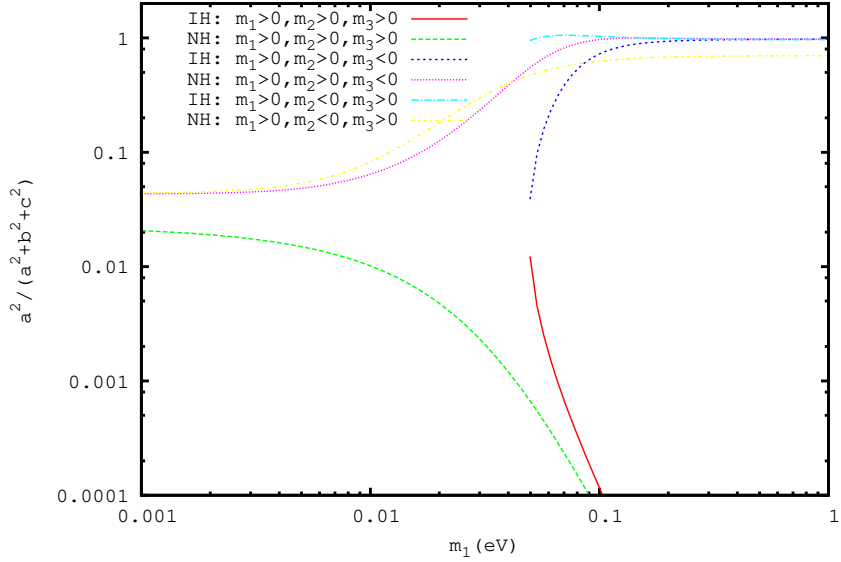}
\caption{We show the plot for the value of $a^2/(a^2+b^2+c^2)$ which correspond to $(y_{\nu}^e)^2$ when we take $\sum_{i=e,\mu\tau}(y_{\nu}^i)^2=1$. The values of $a$, $b$, $c$ are given in  Eq.(3.15)-(3.19) satisfying condition Eq.(3.8). 
We assumed $s_{12}>0$ and $s_{13}>0$ and we fixed $\delta m_{12}^2$, $\delta m_{13}^2$, $s_{12}^2$ and $s_{13}^2$ to the best fit values in Table 1.}
\label{a_param}
\end{figure}

Once we specify the observationally allowed mixing parameters and mass hierarchy where the above one relation is simultaneously satisfied, 
we could determine the all values of neutrino mass model parameters in turn, that is, model parameters of Eq.(3.7) are written by
\begin{eqnarray}
&&a^2=\frac{(m_{\nu})_{12}^2}{(m_{\nu})_{22}}=\frac{(m_{\nu})_{13}^2}{(m_{\nu})_{33}},\\
&&b^2=(m_{\nu})_{22},\\
&&c^2=(m_{\nu})_{33},\\
&&X_A=(m_{\nu})_{11}-a^2=(m_{\nu})_{11}-\frac{(m_{\nu})_{12}^2}{(m_{\nu})_{22}},\\
&&X_B=(m_{\nu})_{23}-bc.
\end{eqnarray}
In Fig.~\ref{a_param}, we show the preferred values for the above model parameter $a$ which may be an important coupling for $\nu_R$ searches in electron-positron colliders when $\eta$ bosons are heavy. 
We find that for $m_1$, $m_2$, $m_3>0$ cases, the coupling takes very small values and it makes the search difficult in the case that only the production of a $\nu_R$ pair is kinematically allowed.

If we do not include $N_4$, $N_5$, $N_6$,  under the assumption of CP invariance in our scalar potential, since radiative corrections obtain universal contributions to $X_A$, $X_B$ except for the neutrino Yukawa coupling dependencies, non-zero $A_4$ symmetric mass matrix elements become
\begin{eqnarray}
(m^{\mbox{sym}})_{11}&=&C_{\mbox{rad}}y_{\nu}^ey_{\nu}^e,\\
(m^{\mbox{sym}})_{23}&=&(m^{\mbox{sym}})_{32}=C_{\mbox{rad}}y_{\nu}^{\mu}y_{\nu}^{\tau}.
\end{eqnarray}

As a result,  another relation has to be imposed, 
\begin{eqnarray}
\frac{(m_{\nu})_{12}}{(m_{\nu})_{22}}\times \frac{(m_{\nu})_{13}}{(m_{\nu})_{33}}=\frac{(m_{\nu})_{11}}{(m_{\nu})_{23}}.
\end{eqnarray}
We find that when we impose the first condition Eq.(3.8), the case where this second condition (3.22) is simultaneously satisfied 
within the $3\sigma$ range of \cite{Globalfit1} does not exist in the case of real $m_1$, $m_2$ and $m_3$. 
In such a case, $N_4$ (and/or $N_5$, $N_6$) is necessarily required to explain the observed neutrino mass structure.

On the other hand, throughout this paper, we have not investigated general cases for CP phases. 
The limited analysis may not obtain the complete information of the prediction of this model. We will present further analysis for general cases of CP phases elsewhere in future.

\section{Right-handed neutrino dark matter}
In this model, as we have seen it in section 2, 
the $A_4$ breaking due to the vacuum alignment 
$(\langle\eta_1\rangle,\langle\eta_2\rangle,\langle\eta_3\rangle)=(v_{\eta},0,0)$ 
leaves a $Z_2$ generator of $A_4$ corresponding to a parity operator unbroken and make the lightest parity odd particle stable 
which can become a viable dark matter candidate. 
In this sense, both of $\eta_i$ and $\nu_R^i$ (i=2,3) can be dark matter candidates. 
In past studies along with $A_4$ discrete dark matter models~\cite{Hirsch:2010ru,Boucenna:2011tj}, 
only the case that $\eta_i$ are dark matters has been considered. 
In this paper, we investigate another case where right-handed neutrino $\nu_R^i$ becomes dark matter 
and pursue the possibility where the thermal freeze-out in early universe obtains the desired relic density required in the current cosmological observations~\cite{Ade:2013zuv}.

To make $\nu_R^i$ stable, the masses ($m_N$) have to be lower than $\eta_i$ masses, and assuming that $\nu_R^i$ obtain the desired relic density through the thermal freeze-out phenomena, 
Yukawa couplings ($y_{\nu}^i$) should be sizable and TeV scale $\eta$ and $\nu_R$ are required. 
In such a situation, to realize the observed small neutrino masses, $v_{\eta}$ at sub MeV range is required as we will discuss the detail below.

In the case of spontaneous $A_4$ breaking taken in the past studies~\cite{Hirsch:2010ru,Boucenna:2011tj}, $\eta$ masses are related to the EW symmetry breaking scale $v_h$ or $v_{\eta}$. 
The smallness of $v_{\eta}$ makes some of scalar particles light and the other modes obtain EW-scale masses, which may make the viable model building difficult for $\nu_R^i$ dark matter scenario. 
Hence we introduce the following soft $A_4$ breaking term to make all modes of $\eta$ heavy, 
\begin{eqnarray}
L^{\mbox{soft}}=-m_{h\eta_1}^2\eta_1^{\dag}h+\mbox{h.c.}.
\end{eqnarray}
This term develops the desired breaking pattern in the $\eta$ VEVs and approximately we find  $v_{\eta}\simeq m_{h\eta_1}^2/m_{\eta}^2\times v_h\sim 0.1\mbox{MeV}$ if $m_{h\eta_1}^2/m_{\eta}^2\sim 10^{-6}$.
Such smallness of the soft term coupling may be realized if the mediation scale of $A_4$ breaking is significantly higher than the $A_4$ breaking scale in hidden sector or if the couplings are non-perturbatively generated. We leave the discussion for future work and just assume the smallness in our following studies.\footnote{ In general,  such a mechanism which introduces the $A_4$ violating soft term may generate other small $A_4$ breaking terms, for an example, yukawa coulings like flavor violating $\bar{L}_e\tau_R \eta_1$ and flavor conserving $\bar{L}_ee_R\eta_1$. On the other hand, for the inclusion of such possible  $A_4$ breaking terms, if the desired $A_4$ breaking pattern is preserved, such couplings are also suppressed $\lesssim 10^{-6}$ as well as the soft term, and  the conclusions in our paper are basically unchanged. If we consider the radiative corrections, such $\eta_1$ number violating dimensionless terms can generate soft term $m_{h\eta_1}^2\eta_1^{\dag}h$ through quantum corrections. When the $A_4$ breaking scale is higher than weak scale and $m_{h\eta_1}\sim O(1\mbox{GeV})$, the $A_4$ violating dimensionless couplings must be $\ll O(10^{-6})$. This may mean that our bilinear term may be induced by radiative corrections and only the term has phenomenological significances for physics we discussed in this paper.} 
By the inclusion of this soft term, the physical spectrum of $\eta$ particles can be independent from the EW symmetry breaking scale $v_h$ and $A_4$ breaking scale $v_{\eta}$. 
We show the physical spectrum of scalar sector in Appendix B. 

Explaining the observed smallness of neutrino masses, we find
\begin{eqnarray}
m_{\nu}\sim 0.1\mbox{eV}
\left(\frac{y_{\nu}}{0.3}\right)^2
\left(\frac{v_{\eta}}{0.1\mbox{MeV}}\right)^2
\left(\frac{1\mbox{TeV}}{m_N}\right).
\end{eqnarray}
Also the observed small neutrino masses require small couplings for $\eta$ number violating couplings
\footnote{This 4-point interaction violates $\eta$ number by $\Delta \eta=2$ and can be independent from the other terms with $\Delta \eta=0$, 1 in the origin.},
$\lambda_{11}\sim O(10^{-8})$ and heavy $N_4$ (and/or $N_5$, $N_6$), $M_{4}\sim 10^{13}\mbox{GeV}$ ($m_{N_5}\sim 10^{13}\mbox{GeV}$) when the Yukawa couplings $Y_i$ ($i=4,5,6$) are of $O(1)$.
\footnote{Very small Yukawa and TeV-scale $m_{N_i}$ ($i=4,5,6$) might be still viable, in this paper, we do not discuss the possibility further more.}  

Next we evaluate the relic density of $\nu_R$ dark matter. 
In this model, the masses of $\nu_R^i$ ($i=1,2,3$) are degenerate at the tree level and the mass splitting arises through loop corrections by picking up $A_4$ breaking $v_\eta$. 
We have to understand the roles of heavier $\nu_R$ state in thermal history. 
The leading contributions for the mass splitting are introduced through following dimension five operators,
\begin{eqnarray} 
&&L_{m_{\eta}}=\frac{M_N}{\Lambda^2}[\eta^{\dag}[\eta[\overline{\nu^c_R} \nu_R]_{\bf 3}]_{\bf 3}]_{\bf 1},\\ 
&&L_{m_{h\eta}}=\frac{M_N}{\Lambda^2}\frac{m_{h\eta_1}^2}{m_{\eta}^2}[\eta^{\dag}h[\overline{\nu_R^c} \nu_R]_{\bf 3}]_{\bf 1},
\end{eqnarray}
where $\Lambda$ is a cut-off scale. It is expected $\Lambda \gg m_{\eta}$ and the mass splitting may be smaller than neutrino mass $m_{\nu}$. 
These terms are proportional to $M_N$ because they would be related to the mechanism to realize TeV scale Majorana mass $M_N$. 
As for parity-even $\nu_R^1$, 
since the decay into $\nu_R^i$ and two leptons is suppressed due to the very small mass splitting, 
it can dominantly decay to SM particles, $\nu_R^1\to h+\nu$ through a mixing between the standard model Higgs and $\eta_1$ bosons,
\begin{eqnarray}
\tau_{\mbox{even}}^{-1}
\sim\frac{y_{\nu}^2}{32\pi}\left(\frac{m_{h\eta_1}^2}{m_{\eta}^2}\right)^2\frac{m_N^2-m_h^2}{m_N}
\sim[10^{-14}\mbox{sec}]^{-1}\left(\frac{y_{\nu}}{1.0}\right)^2\left(\frac{m_{h\eta_1}^2/m_{\eta}^2}{10^{-6}}\right)^2\left(\frac{m_N}{500\mbox{GeV}}\right),
\end{eqnarray}
where $y_{\nu}^2=\sum_{i=e,\mu,\tau} (y^i_{\nu})^2$ is defined.
This means that for $y_{\nu}\sim O(1)$, the parity-even $\nu_R^1$ can be short-lived enough and it may not disturb the thermal relic estimation of $\nu_R^i$ by the late decays. 
In the case for parity-odd $\nu_R$s, 
due to the very tiny mass splitting between heavier and lighter states, 
the decay of the heavier state to the lighter state may be introduced through the transition magnetic moments of $\nu_R^i$,
\begin{eqnarray}
&&L_{\eta}=\frac{c_{\eta}}{\Lambda^3}[\eta^{\dag}[\eta[\overline{\nu^c_R}\sigma^{\mu\nu}\nu_R]_{\bf3}]_{\bf3}]_{\bf1}F_{\mu\nu},\\
&&L_{\eta h}=\frac{c_{\eta h}}{\Lambda^3}\frac{m_{h\eta_1}^2}{m_{\eta}^2}[\eta^{\dag}h[\overline{\nu^c_R}\sigma^{\mu\nu}\nu_R]_{\bf3}]_{\bf1}F_{\mu\nu}.
\end{eqnarray}
Once $A_4$ symmetry is broken by $v_{\eta}$, 
the above interaction generates off diagonal elements and contributes to the decay of heavier state $\nu_R^h$ to lighter state $\nu_R^l$ $\nu_R^h\to\nu_R^l+\gamma$.
\footnote{$L=\frac{c^R_{ij}}{\Lambda}\overline{\nu^c_R}^i\sigma^{\mu\nu}\nu_R^jF_{\mu\nu}$ vanishes because of the Majonara nature of $\nu_R$ and $A_4$ nature.}
The gamma line has very small width and it is very soft $E_{\gamma}<m_{\nu}$. 
We find that the lifetime of the heavier state is longer than the age of the universe, 
\begin{eqnarray}
\tau_{\mbox{odd}}^{-1}\sim \frac{c_{\eta}^2}{64\pi}\frac{v_{\eta}^4}{\Lambda^6}(\delta m_N)^3<\frac{\alpha}{64\pi}\frac{m_{\nu}^5}{m_{\eta}^4}<10^{-23}\tau_U^{-1},
\end{eqnarray}
where $\delta m_N$ is the mass difference between $\nu_R^2$ and $\nu_R^3$ and $\tau_U\sim 13.8\mbox{Gyears}$. 
It may be difficult to detect this line spectrum in CMB at present~\cite{Kim:2011ye}. 
This model realizes multi-state dark matter $\nu_R^2$ and $\nu_R^3$ at present time. 

\begin{figure}[htbp]
\centering\leavevmode
\includegraphics[scale=0.7]{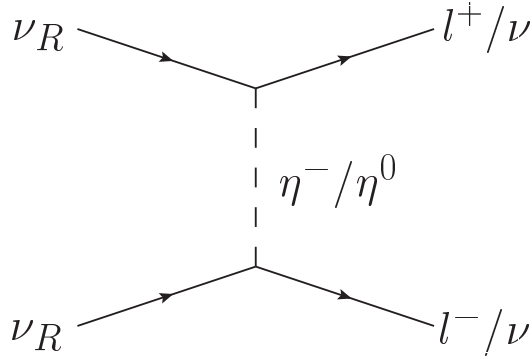}
\caption{The main annihilation processes for $\nu_R$ dark matter during the thermal freeze out.. Since this processes contain two type of majorana fermions ($\nu_R$ and normal light neutrinos), the exchange diagrams among external majorana fermions are included. The dominant piece in NR limit is $O(v_{\mbox{rel}}^2)$.
\label{Annihi}}
\end{figure}

\begin{figure}[htbp]
\centering\leavevmode
\includegraphics[scale=0.5,angle=-90]{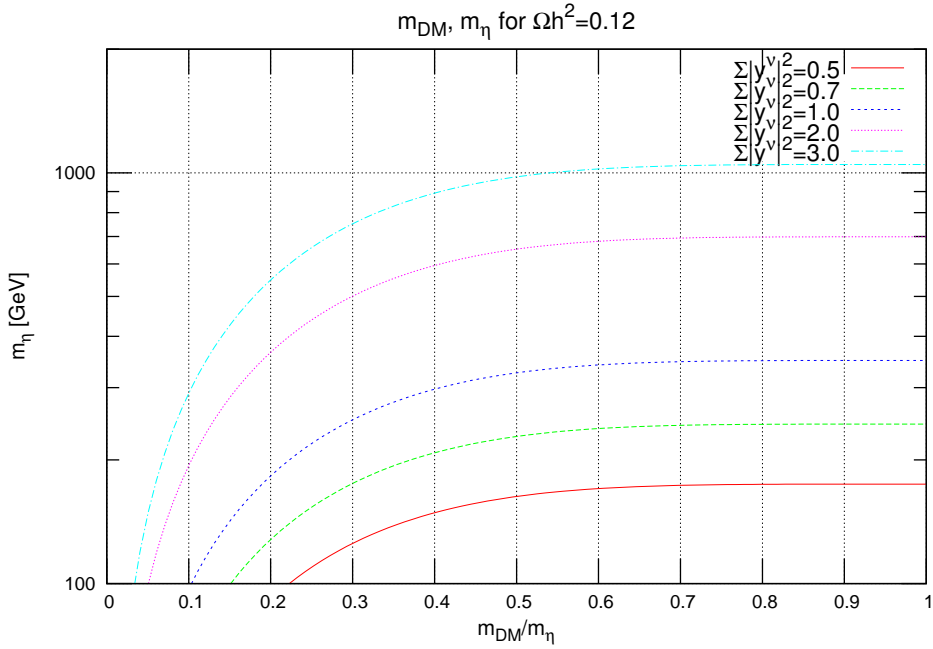}
\caption{The values of $m_{\eta}$ and $m_{N}$ that give the observed relic abundance of dark matter in the case of $\nu_{R}$ dark matter scenario.\label{relic}}
\end{figure}

The main annihilation process of $\nu_R^i$s happens through the process shown in Fig.\ref{Annihi} and the P-wave dominates for the thermal relic estimation. 
The leading term of the annihilation cross section for a single species $\nu_R^2$ (or $\nu_R^3$) is,
\begin{eqnarray}
\sigma_{\mbox{ann}}v_{\mbox{rel}} &\simeq&\frac{y_{\nu}^2}{16\pi}\frac{1}{m_N^2}\frac{1+(m_{\eta}/m_N)^4}{(1+(m_{\eta}/m_N)^2)^4}v_{\mbox{rel}}^2,\nonumber\\
&\sim&
 2.4\mbox{pb}
\left(\frac{v_{\mbox{rel}}^2}{0.3}\right)
\left(\frac{y_{\nu}^{2}}{1.0}\right)^2
\left(\frac{350\mbox{GeV}}{m_N}\right)^2
\left(\frac{(1+(\frac{m_{\eta}}{m_N})^4)/(1+(\frac{m_{\eta}}{m_N})^2)^4}{1/8}\right),
\end{eqnarray}
where $y_{\nu}^2=\sum_{i=e,\mu,\tau}(y_{\nu}^i)^2$, $v_{\mbox{rel}}$ is the relative velocity of incident two dark matter particles. The contributions from higher terms $O(v_{\mbox{ref}}^{2n})$ ($n\geq 2$) give less than 10 percents of the leading contribution in the relic abundance estimation. 
In the thermal relic estimation, we deal with the two states of $\nu_R^i$ as stable. In Fig.\ref{relic}, we show the preferred values of $\eta$, $\nu_R$ masses and neutrino Yukawa coupling to obtain full amount of observed dark matter relic density~\cite{Ade:2013zuv}. 
\footnote{The wrong estimations in Eq(4.9) and Fig.14 in the published version of this paper~\cite{Hamada:2014xha} are corrected. As the result, the prefered mass range for $\eta$ bosons are lowered. Now the constraints from rare lepton decays and EW precision tests may become important since this model obtains radiatively induced $A_4$ symmetric 4-Fermi interactions through one-loop box diagrams~\cite{future}. For the case of $m_{\eta}\simeq m_N$, lepton universality and LEP constraints currently obtain $m_{\eta}\gtrsim (110-140)\mbox{GeV}((y_{\nu}^i)^2/(1/2))$ ($i=e$ or $\mu$ or $\tau$). Rare tau decay $\tau\to\mu\bar{e}e$ imposes $m_{\eta}\gtrsim 130\mbox{GeV}(y_{\nu}^e\sqrt{y_{\nu}^{\mu}y_{\nu}^{\tau}}/(1/5))$ which can be weaken if one of neutrino yukawa couplings is small, e.g in the case of $y_{\nu}^e\ll y_{\nu}^{\mu}\sim y_{\nu}^{\tau}$ allowed by neutrino data as shown in Fig.12.}
Now we understand that in this model, WIMP type dark matter scenario can be achieved by TeV-scale $\nu_R$ and $\eta$, sub MeV $v_{\eta}$ and $O(1)$ neutrino Yukawa couplings.
Here we did not include co-annihilation processes like $\eta_i+\nu_R^i\to l^{\ast}\to l+\mbox{a gauge boson}(W,Z,\gamma)$. 
Such processes are relevant only if the masses of $\nu_R^i$ highly degenerate with those of $\eta_i$.

The collider signals for parity-odd $\eta$ bosons are similar to R-parity conserving minimal supersymmetric standard model(MSSM) with bino dark matter except for the production rate. 
This model has only the pure electroweak productions at the LHC. 
The direct EW production of left-handed sleptons producing multi-lepton final state receives the LHC constraints as 
$m\gtrsim 300\mbox{GeV}$ at ATLAS~\cite{Aad:2014vma} and 
$m\gtrsim 300\mbox{GeV}$ at CMS~\cite{CMS:2013dea} 
depending on the mass splitting of the lightest supersymmetric particle and slepton. 
These constraints include the Drell-Yan production.
\footnote{Gauge boson fusion process also exists. 
The s-channel process is highly suppressed due to the smallness of $v_{\eta}$. Thus, t-channel process is the dominant process, 
but it would be small compared with Drell-Yan processes.} 
We find enough allowed parameter spaces to realize thermal freeze out scenario to obtain desired relic density. 
As for parity-even $\eta_1$,  since $v_{\eta}$ is very small and the di-boson decay mode is suppressed, 
the primary decay is similar to the case of parity-odd $\eta$ bosons though, 
the decay products contain parity-even $\nu_R^1$ decaying to a Higgs and a light neutrino. 
$\nu_R^1$ may be long-lived, which might leave the displaced track in collider detectors. 

Here we consider the possibility for dark matter indirect detections. 
Again the situation is similar to the case of bino dark matter in MSSM, but it differs in the coupling of DM-lepton-$\eta$ which is not fixed by hypercharge gauge coupling. 
Since the dominant $2\to 2$ annihilation process is velocity suppressed or chirality suppressed, 
radiative processes like $\nu_R+\nu_R\to\gamma+l\bar{l}$ may become important~\cite{gamma_spec,Asano:2012zv} 
and the gamma ray signals have the characteristic properties on the spectrum~\cite{gamma_spec}. 
Other indirect detections of these types of dark matter through charged cosmic rays and neutrinos have been intensively studied in past papers~\cite{gamma_pheno}. 

\begin{figure}[htbp]
\centering\leavevmode
\includegraphics[scale=0.7]{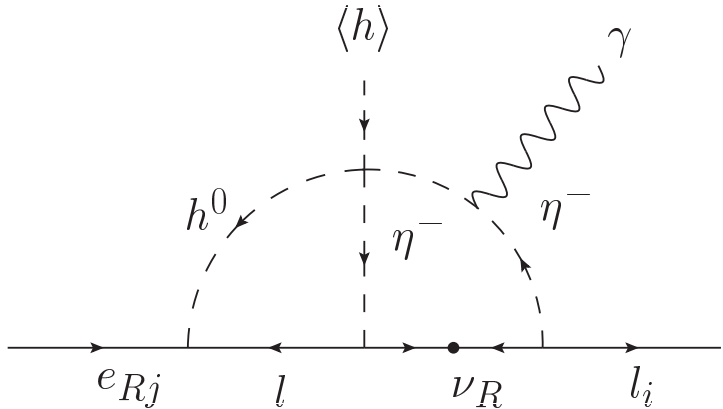}
\caption{A two loop diagram contributing to lepton transition magnetic moments.\label{LFV}}
\end{figure}

In this model, when $v_{\eta}$ goes to zero, $\eta$ and $\nu_R$ couple with only left-handed leptons which do not contribute to the lepton transition magnetic moments at one-loop level.
It is expected that a small contribution arises at two loop level from Fig.~\ref{LFV}. 
The situation is similar in the loop contributions through $N_i$ ($i=4,5,6$) which also do not directly couple with right-handed charged leptons.
\footnote{As we know in MSSM, if we introduce new scalars with the same SM gauge quantum numbers of right-handed sleptons, we expect the sizable contribution to, for example, muon $g-2$. 
However, in this case, the annihilation process of $\nu_{R}^a$ can have S-wave component and may have different implications to the relic density and the indirect detection. 
On the other hand, if the new scalars are $A_4$ charged and do not acquire VEVs, LFV may be suppressed by $A_4$ symmetric nature as we will see below.}
Such loop contributions may be described by the following dimension six operator,
\begin{eqnarray}
L=\frac{c_{ij}}{\Lambda^2}\overline{L}_ih\sigma^{\mu\nu}(e_R)_jF_{\mu\nu},
\end{eqnarray} 
where $c_{ij}$ are $O(1)$ numerical coefficients, $\Lambda$ is a cut off scale of effective operators and it is expected to be higher than $\eta$ mass scale.  
We might expect the lepton flavor violating(LFV) contributions like $\mu\to e\gamma$ , $\tau\to e\gamma$ due to the dimension six operator, 
however, when we ignore the $v_{\eta}/v_h$, this term can not have LFV contributions due to the conservation of $Z_3$ charge of $A_4$ and only flavor diagonal contributions like muon $g-2$ may be allowed.  
The non-zero LFV contributions through lepton transition magnetic moments require the $Z_3$ symmetry violation, that is, the $\eta$ VEV. 
They can arise at one-loop level through mediators, $\nu_R$ and $N_i$ ($i=4,5,6$)), but they face the significant suppression due to the small $v_\eta/\Lambda\ll 1$. 
The LFV process with no chirality flips through Z boson couplings is also aligned to diagonal form 
due to $A_4$ symmetry nature $Y_{\nu}Y_{\nu}^{\dag}=3{\rm diag}((y_{\nu}^e)^2,(y_{\nu}^{\mu})^2,(y_{\nu}^{\tau})^2)$ if $v_{\eta}$ is not picked up, 
and it is suppressed again as well as the case of the magnetic moment type LFV processes.  
In this model, $A_4$ symmetry remains as an approximately good symmetry at low energy and it plays a key role to suppress LFV processes in nature. \footnote{Even though we add  other explicit $A_4$ breaking terms, this statement may be correct as long as the couplings of added $A_4$ breaking are small.}
This is a contrast to the case of only very heavy right-handed Majorana neutrinos added to Standard Model particles where such flavor symmetry may not necessarily play any role to explain the tiny LFV.

At the last of this section, we consider constraints on $N_i$ ($i=4,5,6$). 
The mass scale of these particles are rather free and only the combination of the masses and neutrino Yukawa couplings $Y_i$ ($i=4,5,6$) are constrained by neutrino masses as the case of usual seesaw mechanism. 
If we assume TeV-scale $N_i$ ($i=4,5,6$), the lifetime is $O(1)\times 10^{-14}\mbox{sec}\times (m_{N_i}/1\mbox{TeV})^{-1}$. 
This production at collider may be minor if the mass is heavier than SM Higgs mass. 
On the other hand, in early universe, it may play some roles at the freeze out time of dark matter or the later, e.g. diluting dark matter relic at late time. 
Thus we simply assume that they have heavy masses, for example, $\sim 10^{12-13}\mbox{GeV}$.

\section{Conclusion and discussion}
In this paper, we discussed the discrete dark matter model originally introduced in \cite{Hirsch:2010ru,Boucenna:2011tj} 
and showed that this type of models can explain current experimental results of neutrino masses and mixing angles, 
that is, it can achieve non-zero $\theta_{13}$. 
We find that this model predicts one relation among neutrino mass matrix elements and the non-zero $\theta_{13}$ requires non-zero $\delta s_{23}$ and $m_1$ in both NH and IH cases assuming no CP phases. 
Such prediction can be tested in several future neutrino experiments and cosmological observations. 
Next, we investigated the possibility of $\nu_R$ dark matter, especially focusing on the case that they obtain the desired relic density of observed dark mater. 
This motivates the existence of TeV-scale $\nu_R$. We could realize such a possibility by introducing an explicit  $A_4$ breaking bilinear term. 
We find that the current experimental constraints still allow the scenario that the thermal freeze out of $\nu_R$ dark matter obtains the desired relic density. 
Future collider experiments such as the LHC and the ILC may discover the signals or exclude the large parts of interesting parameter spaces. 
Within TeV-scale $\nu_R$ scenario, the $A_4$ symmetry plays an interesting role to hide LFV processes in low energy physics. 
We demonstrated that even the two loop processes can be hidden due to the symmetry, 
and LFV processes only appear when the breaking is picked up, which is highly suppressed by the mismatch of $v_{\eta}$ and cut off scale $\gtrsim m_{\eta}$. 
This is a contrast to heavy right-handed neutrino scenarios in the role of flavor symmetry.

In this paper, we only considered the possibility of TeV-scale $\nu_R$ though, 
notice that the TeV-scale mass is required when we assume that the thermal freeze out obtain the desired relic density of the present dark matter. 
The physical mass of $\eta$ is not related to the EW symmetry breaking scale any more thanks to the soft term $m_{h\eta_1}^2h^{\dag}\eta$ and the $A_4$ symmetric $\eta$ mass term $m_{\eta}^2\eta^{\dag}\eta$. 
Even in the case that $\nu_R$ significantly heavier than 1TeV, 
we could realize that the $\eta$ is heavier than $\nu_R$. 
If we relax the constraints on thermal freeze out for the observed dark matter density, heavy $\nu_R$, e.g stable $10^{12}$GeV right-handed neutrinos may be allowed 
and may obtain other possibilities within the $\nu_R$ dark matter scenario, e.g. possibilities of the simultaneous production of dark matter and baryon asymmetry, which was not discussed in this paper. 
For such heavy $\nu_R$, $m_{h\eta}^2/m_{\eta}^2$ is not necessarily very small and the small neutrino masses are achieved in the usual meaning of Type I seesaw mechanism. 
On the other hand, the EW symmetry breaking may require a fine tuning among $m_h^2$, $m_{h\eta}^2,$ and $m_{\eta}^2$ at the EW scale, 
which may be theoretical challenges in different points of view from the case of TeV-scale $\nu_R$.
\footnote{See the scalar boson spectrum and the condition for the EW symmetry breaking in Appendix A.} 
The most of our phenomenological discussions presented in this paper depend on only $v_{\eta}/\Lambda$. 
By fixing the ratio, we may find similar conclusions except for the testability in collider experiments.

\subsection*{Acknowledgement}
This work is supported in part by the Grant-in-Aid for Scientific Research No. 25400252 (T.~K.), No. 23104011 (Y.~O.), No.25.1146 (A.~O.), No25.1107 (Y.~H.) from the Ministry of Education, Culture, Sports, Science and Technology of Japan. 

\appendix

\section{A short glance at $A_{4}$ group}

$A_{4}$ is the group of even permutation of four objects. 
In this appendix, we show some properties of $A_{4}$ which is needed to describe the discrete dark matter model. 

$A_{4}$ has four irreducible representations ${\bf 1,1',1'',3}$, and is generated by two generators $S, T$ which satisfy
\begin{eqnarray}
S^{2}=T^{3}=1, \quad (ST)^{3}=1.
\end{eqnarray}
On the trivial singlet {\bf 1}, $S$ and $T$ are represented by $S=1$ and $T=1$. ${\bf 1'}({\bf 1''})$ corresponds to $S=1,T=\omega(\omega^{2})$. Here, $\omega$ is a primitive cube root of 1, say $e^{2\pi i/3}$ . 
On {\bf 3} representation, $S$ and $T$ is represented by
\begin{eqnarray}
S=\left(\begin{array}{ccc}
1 &  0 &  0 \\
0 & -1 &  0 \\
0 &  0 & -1
\end{array}\right), \quad
T=\left(\begin{array}{ccc}
0 & 1 & 0 \\
0 & 0 & 1 \\
1 & 0 & 0
\end{array}\right).
\end{eqnarray}
The sub group of $A_{4}$ generated by $S$ is left as the symmetry of the discrete dark matter model even after the scalar fields get VEVs. This subgroup $Z_{2}$ guarantees stability of a dark matter candidate. 
Multiplication rule is as below,
\begin{eqnarray}
&&{\bf 1' \otimes 1' = 1'', \quad 1'' \otimes 1'' = 1', \quad 1' \otimes 1'' = 1,}\nonumber\\
&&{\bf 3 \otimes 3 = 3_{\rm 1} \oplus 3_{\rm 2} \oplus 1 \oplus 1' \oplus 1''.}
\end{eqnarray}

For example, when $a=(a_{1},a_{2},a_{3})$ and $b=(b_{1},b_{2},b_{3})$ are two $A_{4}$ triplets, the ways to compose ${\bf 1, 1', 1''}$ and {\bf 3} representation from them are
\begin{eqnarray}
&&(ab)_{\bf 1}=a_{1}b_{1}+a_{2}b_{2}+a_{3}b_{3},\nonumber\\
&&(ab)_{\bf 1'}=a_{1}b_{1}+\omega a_{2}b_{2}+\omega^{2}a_{3}b_{3},\nonumber\\
&&(ab)_{\bf 1''}=a_{1}b_{1}+\omega^{2}a_{2}b_{2}+\omega a_{3}b_{3},\nonumber\\
&&(ab)_{\bf 3_{\rm 1}}=\left(\begin{array}{c}
                           a_{2}b_{3} \\
                           a_{3}b_{1} \\
                           a_{1}b_{2} 
                           \end{array}\right), \quad
  (ab)_{\bf 3_{\rm 2}}=\left(\begin{array}{c}
                           a_{3}b_{2} \\
                           a_{1}b_{3} \\
                           a_{2}b_{1} 
                           \end{array}\right).
\end{eqnarray}

\section{Scalar boson potential and the physical spectrum}
General form of CP and $A_4$ invariant potential terms of scalar bosons are given by,
\begin{eqnarray}
V(h,\eta)&=&m_{\eta}^{2}\eta^{\dagger}\eta+m_{h}^{2}h^{\dagger}h\nonumber\\
         &&+\lambda_{1}(h^{\dagger}h)^{2}+\lambda_{2}[\eta^{\dagger}\eta]_{1}^{2}+\lambda_{3}[\eta^{\dagger}\eta]_{1'}[\eta^{\dagger}\eta]_{1''}\nonumber\\
         &&+\lambda_{4}([\eta^{\dagger}\eta^{\dagger}]_{1'}[\eta\eta]_{1''}+[\eta^{\dagger}\eta^{\dagger}]_{1''}[\eta\eta]_{1'})+\lambda_{5}[\eta^{\dagger}\eta^{\dagger}]_{1}[\eta\eta]_{1}\nonumber\\
         &&+\lambda_{6}([\eta^{\dagger}\eta]_{3_{1}}[\eta^{\dagger}\eta]_{3_{1}}+[\eta^{\dagger}\eta]_{3_{2}}[\eta^{\dagger}\eta]_{3_{2}})+\lambda_{7}[\eta^{\dagger}\eta]_{3_{1}}[\eta^{\dagger}\eta]_{3_{2}}+\lambda_{8}[\eta^{\dagger}\eta^{\dagger}]_{3_{1}}[\eta\eta]_{3_{1}}\nonumber\\
         &&+\lambda_{9}[\eta^{\dagger}\eta]_{1}(h^{\dagger}h)+\lambda_{10}[\eta^{\dagger}h]_{3}[h^{\dagger}\eta]_{3}+\lambda_{11}([\eta^{\dagger}\eta^{\dagger}]_{1}hh+h^{\dagger}h^{\dagger}[\eta\eta]_{1})\nonumber\\
         &&+\lambda_{12}([\eta^{\dagger}\eta^{\dagger}]_{3_{1}}[\eta h]_{3}+[h^{\dagger}\eta^{\dagger}]_{3}[\eta\eta]_{3_{2}})+\lambda_{13}([\eta^{\dagger}\eta^{\dagger}]_{3_{2}}[\eta h]_{3}+[h^{\dagger}\eta^{\dagger}]_{3}[\eta\eta]_{3_{1}})\nonumber\\
         &&+\lambda_{14}([\eta^{\dagger}\eta]_{3_{1}}[\eta^{\dagger}h]_{3}+[h^{\dagger}\eta]_{3}[\eta^{\dagger}\eta]_{3_{2}})+\lambda_{15}([\eta^{\dagger}\eta]_{3_{2}}[\eta^{\dagger}h]_{3}+[h^{\dagger}\eta]_{3}[\eta^{\dagger}\eta]_{3_{1}}).\nonumber\\
\end{eqnarray}
To explain observed tiny neutrino masses in our scenario, we have to demand smallness for $m_{h\eta_1}^2$ and $\lambda_{11}$. The quantum corrections due to 
$\Delta \eta=1$ interactions, $\lambda_{12}$, $\lambda_{13}$, $\lambda_{14}$ and $\lambda_{15}$ generate $\lambda_{11}$ at one loop, so these conpligs also have to be 
suppressed $<m_{h\eta_1}^2/m_{\eta}^2$. This may exhibit an approximate global $U(1)_{\eta}$
symmetry in the scalar potential. Notice that $\lambda_{11}$ also violates $U(1)_{\eta}$ by $\Delta \eta=2$ but the quantum corrections by itself never generate $\Delta \eta=1$ interactions.

As we mentioned in section 2, we add the following $A_4$ explicit breaking term, 
\begin{eqnarray}
V_{\mbox{soft}}=-m_{h\eta_1}^2\eta_1^{\dag}h+\rm{h.c.},
\end{eqnarray}
which explicitly breaks $U(1)_{\eta}$ by $\Delta \eta=1$. 

We notice that in this scalar potential, an exact invariance for an odd permutation between $\eta_2$ and $\eta_3$ exists.  The full invariance for all three odd permutations among $\eta_1$, $\eta_2$ and $\eta_3$ recovers if we ignore the soft term $m_{h\eta_1}^2$. The $(\eta_2,\eta_3)$ permutation is not a symmetry inside $A_4$ symmetry but an accidental symmetry in our model when we impose CP invariance in scalar potential. As we explain in Appendix D, this invariance for $(\eta_2,\eta_3)$ permutation is crucial to obtain a relation of Eq.(3.8) in neutrino mass matrix elements. CP invariance in all couplings of the scalar potential is not always nesessary for the invariance of ($\eta_2,\eta_3$) odd permutation in scalar potential, for example, the CP invariance in $\lambda_{11}$ coupling can be relaxed for this purpose. The phase of $\lambda_{11}$ can introduce CP phases for neutrino mass matrix without changing the relation Eq.(3.8). In general, inclusions of CP phases in the other terms of scalar potential may violate the invariance for ($\eta_2, \eta_3$) permutation, for example, by the following term,
\begin{eqnarray}
\lambda_4[\eta^{\dag}\eta^{\dag}]_{1'}[\eta\eta]_{1"}+
\lambda_{4'}[\eta^{\dag}\eta^{\dag}]_{1"}[\eta\eta]_{1'}, 
\end{eqnarray}
where $\lambda_4\neq\lambda_{4'}$. In such cases, the relation Eq.(3.8) is not hold any more, which results more freedom to describe neutrino mass matrix in this model.

We expand the fields around the physical vacuum $\langle h\rangle=v_h$, $(\langle\eta_1\rangle,\langle\eta_2\rangle,\langle\eta_3\rangle)=(v_{\eta},0,0)$,
\begin{eqnarray}
h=\left(\begin{array}{c}
h^+\\
v_h+h^0+iA^0_h
\end{array}
\right)
,~
\eta_1=\left(\begin{array}{c}
\eta_1^+\\
v_{\eta}+\eta_1^0+iA_{\eta_1}^0
\end{array}
\right)
,~
\eta_{2,3=}
\left(\begin{array}{c}
\eta_{2,3}^+\\
\eta^0_{2,3}+iA_{\eta_{2,3}}^0
\end{array}
\right).
\end{eqnarray}
We define new couplings as follows~\cite{Boucenna:2011tj}, 
\begin{eqnarray}
&&L=\lambda_9+\lambda_{10}+2\lambda_{11},\\
&&Q=\lambda_{12}+\lambda_{13}+\lambda_{14}+\lambda_{15},\\
&&P=\lambda_2+\lambda_3+2\lambda_4+\lambda_5,\\
&&R_1=-3\lambda_3-6\lambda_4+2\lambda_6+\lambda_7+\lambda_8,\\
&&R_2=-3\lambda_3-2\lambda_4-4\lambda_5-2\lambda_6+\lambda_7+\lambda_8,\\
&&R_3=-3\lambda_3-4\lambda_4-2\lambda_5+\lambda_8.
\end{eqnarray}
Then the minimalization conditions for scalar potential are written by,
\begin{eqnarray}
&&m_h^2+2\lambda_1v_h^2+Lv_{\eta}^2-m_{h\eta}^2\frac{v_{\eta}}{v_h}=0,\\
&&m_{\eta}^2+2Pv_{\eta}^2+Lv_h^2-m_{h\eta}^2\frac{v_h}{v_{\eta}}=0.
\end{eqnarray}
From the second condition, we approximately read $v_{\eta}\sim \frac{m_{h\eta_1}^2}{m_{\eta}^2}v_h$ when $m_{h\eta_1}^2/m_{\eta}^2\ll 1$.

\subsection{Physical spectrum of scalar bosons}
The physical states of $Z_2$ even and parity even charged Higgs boson sector are
\begin{eqnarray}
h_0^+=\frac{v_h}{v}h^+-\frac{v_{\eta}}{v}\eta^+,~h_1^+=\frac{v_{\eta}}{v}h^++\frac{v_h}{v}\eta^+,
\end{eqnarray}
where $v=\sqrt{v_h^2+v_{\eta}^2}$.
The physical mass spectrum is obtained as
\begin{eqnarray}
m_{h_0^+}^2=0,~m_{h_1^+}^2=(\frac{m_{h\eta}^2}{v_h v_{\eta}}-\lambda_{10}-\lambda_{11})v^2.
\end{eqnarray}

The physical states of $Z_2$ even and parity even neutral Higgs boson sector are,
\begin{eqnarray}
h_0^0=h^0\cos\phi-\eta_1^0\sin\phi,~h_1^0=h^0\sin\phi+\eta_1^0\cos\phi,
\end{eqnarray}
where the mixing angle is
\begin{eqnarray}
\tan 2\phi=\frac{2Lv_{\eta}v_h+m_{h\eta}^2}{2Pv_{\eta}^2-2\lambda_1v_h^2-\frac{m_{h\eta}^2}{2}\left(\frac{v_h}{v_{\eta}}-\frac{v_{\eta}}{v_h}\right)},
\end{eqnarray}
and the mass spectrum is written by
\begin{eqnarray}
m_{h_{0,1}}^2&=&2\lambda_1v_h^2+2Pv_{\eta}^2+\frac{m_{h\eta}^2}{2v_hv_{\eta}}v^2\nonumber\\
&&~~\pm \sqrt{\left(2\lambda_1v_h^2-2Pv_{\eta}^2-\frac{m_{h\eta}^2}{2v_hv_{\eta}}(v_h^2-v_{\eta}^2)\right)^2+\left(2L-\frac{m_{h\eta}^2}{v_hv_{\eta}}\right)^2v_h^2v_{\eta}^2}.
\end{eqnarray}

The physical states of $Z_2$ even and parity odd neutral pseudo scalar Higgs boson sector are
\begin{eqnarray}
A_0^0=\frac{v_h}{v}A_h^0-\frac{v_{\eta}}{v}A_{\eta_1}^0,~A_1^0=\frac{v_{\eta}}{v}A_h^0+\frac{v_h}{v}A_{\eta_1}^0,
\end{eqnarray} 
and the mass spectrum is written by
\begin{eqnarray}
m_{A_0}^2=0,~m_{A_1}^2=(\frac{m_{h\eta}^2}{v_hv_{\eta}}-4\lambda_{11})v^2.
\end{eqnarray}

The physical states of $Z_2$ odd and parity even charged Higgs boson sector are,
\begin{eqnarray}
h_2^+=\frac{1}{\sqrt{2}}(\eta_2^+-\eta_3^+),~h_3^+=\frac{1}{\sqrt{2}}(\eta_2^++\eta_3^+),
\end{eqnarray}
and the mass spectrum is written by,
\begin{eqnarray}
m_{h_{2,3}^+}^2=R_3v_{\eta}^2-(\lambda_{10}+2\lambda_{11})v_h^2+m_{h\eta}^2\frac{v_h}{v_{\eta}}\pm Qv_hv_{\eta}.
\end{eqnarray}

The physical states of $Z_2$ odd and parity even neutral Higgs boson are,
\begin{eqnarray}
h_2^0=\frac{1}{\sqrt{2}}(\eta_2^0-\eta_3^0),~h_3^0=\frac{1}{\sqrt{2}}(\eta_2^0+\eta_3^0),
\end{eqnarray}
and the mass spectrum is written by,
\begin{eqnarray}
m_{h_{2,3}}^2=R_1v_{\eta}^2+m_{h\eta}^2\frac{v_h}{v_{\eta}}\pm Q v_h v_{\eta}.
\end{eqnarray}

The physical sates of $Z_2$ odd and parity odd neutral Higgs boson are,
\begin{eqnarray}
A_2^0=\frac{1}{\sqrt{2}}(A_{\eta_2}^0-A_{\eta_3}^0),~A_3^0=\frac{1}{\sqrt{2}}(A_{\eta_2}^0+A_{\eta_3}^0),
\end{eqnarray}
and the mass spectrum is written by,
\begin{eqnarray}
m_{A_{2,3}}^2=R_2v_{\eta}^2-4\lambda_{11}v_h^2+m_{h\eta}^2\frac{v_h}{v_{\eta}}\pm Q v_hv_{\eta}.
\end{eqnarray}

We find that the zero mass states are absorbed into the longitudinal components of electroweak massive gauge bosons. 

\section{Neutrino mass matrix and the prediction of discrete dark matter models}
Using the conventional form for PMNS matrix in Eq.~(3.10), we can relate the neutrino mass matrix elements to the observed masses and mixing parameters as follow,
\begin{eqnarray}
(m_{\nu})=
U_{\mbox{PMNS}}
\left(\begin{array}{ccc}
|m_1|&0&0\\
0&|m_2|&0\\
0&0&|m_3|
\end{array}\right)
U^T_{\mbox{PMNS}}
=
\left(\begin{array}{ccc}
(m_{\nu})_{11}&(m_{\nu})_{12}&(m_{\nu})_{13}\\
(m_{\nu})_{12}^{\ast}&(m_{\nu})_{22}&(m_{\nu})_{23}\\
(m_{\nu})_{13}^{\ast}&(m_{\nu})_{23}^{\ast}&(m_{\nu})_{33}
\end{array}\right),
\end{eqnarray}
\begin{eqnarray}
(m_{\nu})_{11}&=&c_{13}^2(m_1c_{12}^2+s_{12}^2m_2)+s_{13}^2m_3,\\
(m_{\nu})_{22}&=&-2s_{12}c_{12}s_{23}c_{23}s_{13}\delta m_{12}\cos\delta\nonumber\\
&&~~~~~+c_{23}^2(s_{12}^2m_1+c_{12}^2m_2)+s_{23}^2s_{13}^2(c_{12}^2m_1+s_{12}^2m_2)+
s_{23}^2c_{13}^2m_3,\\
(m_{\nu})_{33}&=&2s_{12}c_{12}s_{23}c_{23}s_{13}\delta m_{12}\cos\delta\nonumber\\
&&~~~~~+s_{23}^2(s_{12}^2m_1+c_{12}^2m_2)+c_{23}^2s_{13}^2(c_{12}^2m_1+s_{12}^2m_2)+
c_{23}^2c_{13}^2m_3,\\
(m_{\nu})_{12}&=&s_{12}c_{12}c_{23}c_{13}\delta m_{12}-s_{23}s_{13}c_{13}e^{-i\delta}(c_{12}^2m_1+s_{12}^2m_2-m_3),\\
(m_{\nu})_{13}&=&-s_{12}c_{12}s_{23}c_{13}\delta m_{12}-c_{23}s_{13}c_{13}e^{-i\delta}(c_{12}^2m_1+s_{12}^2m_2-m_3),\\
(m_{\nu})_{23}&=&s_{12}c_{12}s_{13}(s_{23}^2e^{i\delta}-c_{23}^2e^{-i\delta})\delta m_{12}
\nonumber\\
&&~~~~~
-s_{23}c_{23}\left((s_{12}^2m_1+c_{12}^2m_2)-s_{13}^2(c_{12}^2m_1+c_{12}^2m_2)
\right)+s_{23}c_{23}c_{13}^2m_3,\nonumber\\
\end{eqnarray}
where the masses are defined as $m_1=|m_1|$, $m_2=|m_2|e^{i\phi_2}$ and $m_3=|m_3|e^{i\phi_3}$.

We can find the following structure for $(m_{\nu})_{12}$, $(m_{\nu})_{13}$, $(m_{\nu})_{22}$ and $(m_{\nu})_{33}$,
\begin{eqnarray}
&&(m_{\nu})_{22}=-As_{23}c_{23}+Bc_{23}^2+Cs_{23}^2, \\
&&(m_{\nu})_{33}=As_{23}c_{23}+Bs_{23}^2+Cc_{23}^2,\\
&&(m_{\nu})_{12}=Xc_{23}-Ys_{23},\\
&&(m_{\nu})_{13}=-Xs_{23}-Yc_{23},
\end{eqnarray}
\begin{eqnarray}
&&A=2s_{12}c_{12}s_{13}\delta m_{12}\cos\delta,\\
&&B=(s_{12}^2m_1+c_{12}^2m_2),\\
&&C=s_{13}^2(c_{12}^2m_1+s_{12}^2m_2)+c_{13}^2m_3,\\
&&X=s_{12}c_{12}c_{13}\delta m_{12},\\
&&Y=s_{13}c_{13}e^{-i\delta}((c_{12}^2m_1+s_{12}^2m_2)-m_3).
\end{eqnarray}
Our discrete dark matter model predicts Eq.(3.8). The condition $(m_{\nu})_{12}^2/(m_{\nu})_{22}=(m_{\nu})_{13}^2/(m_{\nu})_{33}$ gives
\begin{eqnarray}
&&\frac{1}{\tan(2\theta_{23})}=\frac{1}{2}\times \frac{A(X^2+Y^2)-2(B+C)XY}{BY^2-CX^2},\\
&&s_{23}=\sin(\frac{\arctan(2\theta_{23})}{2}).
\end{eqnarray}
Notice that $A\propto \delta m_{12} s_{13}s_{12}$, $X\propto \delta m_{12}s_{12}$ and $Y\propto s_{13}$. Then we find that the right hand side of Eq.(C.17) is,
\begin{eqnarray}
\frac{A(X^2+Y^2)-2(B+C)XY}{BY^2-CX^2}\propto \delta m_{12} s_{13}s_{12}.
\end{eqnarray}
If we take $s_{13}=0$, then $c_{23}^2=s_{23}^2=1/2$ is required and Tri-bimaximal mass pattern taken in original paper\cite{Hirsch:2010ru,Boucenna:2011tj} can be realized. 
On the other hand, if we take non-zero $s_{13}$, in general, $\delta s_{23}$ is proportional to $\delta m_{12}$. 
Since the observed $\delta m_{12}$ is not zero, we can expect non-zero deviation $\delta s_{23}$ from $s_{23}^2=1/2$.

For $m_i>0$ ($i=1,2,3$) and degenerate spectrum, we obtain $\delta m_{12}\ll m_1$. 
Then we obtain $CX^2\ll BY^2$ even in the case of small $s_{13}\sim 0.15$. 
In such a case, we find the approximate formula presented as Eq.(3.14). 

In the case of real $m_1$, $m_2$ and $m_3$, the imaginary part of the condition Eq.(3.9) obtain
\begin{eqnarray}
&&([-A(c_{12}^2m_1+s_{12}^2m_2-m_3)s_{13}c_{13}\cos\delta+(B+C)X]s_{23}c_{23}\nonumber\\
&&~~~~~+B(c_{12}^2m_1+s_{12}^2m_2-m_3)(c_{23}^2-s_{23}^2)s_{13}c_{13}\cos\delta)\sin\delta=0.
\end{eqnarray}
The possible choice is $\sin\delta=0$, that is, $\delta=0,\pi$.

Taking $s_{23}=\mbox{sgn}(s_{23})/\sqrt{2}+\delta s_{23}$, for the cases of no CP phases, the following relation among neutrino masses and mixing parameters is derived, 
\begin{eqnarray}
\delta s_{23}
=\frac{(s_{12}s_{13}/2\sqrt{2})\delta m_{12}(m_1m_2-(c_{12}^2-s_{12}^2)\delta m_{12} m_3 -m_3^2)}{\left(s_{12}^2c_{12}^2[s_{13}^2(c_{12}^2m_1+s_{12}^2m_2)+c_{13}^2m_3](\delta m_{12})^2
-s_{13}^2(s_{12}^2m_1+c_{12}^2m_2)(c_{12}^2m_1+s_{12}^2m_2-m_3)^2\right)}.\nonumber\\
\end{eqnarray}

\section{Radiatively induced neutrino mass structure in discrete dark matter model}
As we mentioned in section 3, the model can generate neutrino masses through radiative correction at loop level. 
Here we explain that the radiative corrections induce the mass structure described in section 3. 
Since $\eta$ boson is almost diagonal in mass, here we take mass insertion approximation.

At one loop, the flavor mixing of $\nu_R$s is highly suppressed at the order of $(v_{\eta}/m_{\eta})^2$ or higher order. 
This requires that the $\eta$ boson propagating inside the loop diagram can not change the flavor indices to connect with internal $\nu_R$ line. 
Another restriction comes from the special pattern of $\eta$ VEVs $(\langle\eta_1\rangle,\langle\eta_2\rangle,\langle\eta_3\rangle)=(v_{\eta}, 0,0)$. 

The first type arises through $\lambda_{11}$ coupling(See Fig.3.). In this case, the $\eta$ propagating internal line of the loop has universal couplings for all the indices $(i=1,2,3)$ in the four points scalar interactions. 
The $SU(2)_L$ breaking in neutrino masses happens through two $v_h$ and there is no $A_4$ breaking part in this diagram at the leading piece. 
This type obtains $A_4$ symmetric mass structure $m^{\mbox{sym}}$. 
That is, the non zero pieces are  
\begin{eqnarray}
&&(m_{\nu})^{\mbox{rad:1}}_{11}=\lambda_{11}y_{\nu}^{e}y_{\nu}^{e}\frac{v_h^2}{m_N}f(m_{\eta}, m_N),\\
&&(m_{\nu})^{\mbox{rad:1}}_{23}=(m_{\nu})^{\mbox{rad:1}}_{32}=\lambda_{11}y_{\nu}^{\mu}y_{\nu}^{\tau}\frac{v_h^2}{m_N}f(m_{\eta}, m_N),
\end{eqnarray}
where $f(x,y)\sim\frac{1}{16\pi^2}\frac{y^2}{x^2-y^2}[\frac{y^2}{x^2-y^2}][\log(\frac{x^2}{y^2})+1]$ is a loop function. 
We used $\lambda_{11}v_{\eta}^2\ll (m_{\eta}^2-m_N^2)$.

The second type arises through $\lambda_{4}$ and $\lambda_{5}$ couplings(See Fig.1.). 
In this case, both two of $\eta$s acquire the VEV and the two $\eta$ bosons constitute singlets ($1,1',1''$), 
we find that the common piece of $\eta_2$ and $\eta_3$ loop vanishes due to $Z_3$ nature $1+\omega+\omega^2=0$ 
and the mismatch of the two couplings, $\lambda_{5}-\lambda_{4}$ allows the non zero contributions for  $\eta_2$ and $\eta_3$ loops. 
As a result, we find two mass structures. 
The first one is $A_4$ symmetric mass structure $m^{\mbox{sym}}$ which is proportinal to $\lambda_5-\lambda_4$,  
\begin{eqnarray}
&&(m_{\nu})^{\mbox{rad:2,sym}}_{11}=(\lambda_{5}-\lambda_{4})y_{\nu}^{e}y_{\nu}^{e}\frac{v_{\eta}^2}{m_N}f(m_{\eta}, m_N),\\
&&(m_{\nu})^{\mbox{rad:2,sym}}_{23}=(m_{\nu})^{\mbox{rad:1}}_{32}=(\lambda_{5}-\lambda_{4})y_{\nu}^{\mu}y_{\nu}^{\tau}\frac{v_{\eta}^2}{m_N}f(m_{\eta}, m_N),
\end{eqnarray}
and the second one is $A_4$ violating mass structure $m^{\mbox{break}}$ which is proportional to 3$\lambda_4$,
\begin{eqnarray}
(m_{\nu})^{\mbox{rad:2,break}}_{ij}=3\lambda_4y_{\nu}^iy_{\nu}^j\frac{v_{\eta}^2}{m_N}f(m_{\eta},m_N).
\end{eqnarray}

The third type arises through $\lambda_2$ and $\lambda_3$ couplings. 
In this case, the $\eta$ constitute singlets with the $\eta$ propagating internal line in the loop. 
Since only $\eta_1$ acquire non zero VEV, only $\eta_1$ can be allowed to propagate the internal line 
which results the same mass pattern given in the type I tree level seesaw contribution, 
that is, $A_4$ violating mass structure, $m^{\mbox{break}}$, 
\begin{eqnarray}
(m_{\nu})^{\mbox{rad:3,break}}_{ij}=3(\lambda_2+\lambda_3)y_{\nu}^iy_{\nu}^j\frac{v_{\eta}^2}{m_N}f(m_{\eta},m_N).
\end{eqnarray}

The forth type is induced through $\lambda_6$ coupling. In this case, only $\eta_2$ and $\eta_3$ are allowed to propagate the internal line of loops. Then we find that this contribution is regarded as the sum of $m^{\mbox{sym}}$,  
\begin{eqnarray}
&&(m_{\nu})^{\mbox{rad:4,sym}}_{11}=\lambda_{6}y_{\nu}^{e}y_{\nu}^{e}\frac{v_{\eta}^2}{m_N}f(m_{\eta}, m_N),\\
&&(m_{\nu})^{\mbox{rad:4,sym}}_{23}=(m_{\nu})^{\mbox{rad:4}}_{32}=\lambda_{6}y_{\nu}^{\mu}y_{\nu}^{\tau}\frac{v_{\eta}^2}{m_N}f(m_{\eta}, m_N),
\end{eqnarray}
and $m^{\mbox{break}}$,
\begin{eqnarray}
(m_{\nu})^{\mbox{rad:4,break}}_{ij}=-\lambda_{6}y_{\nu}^iy_{\nu}^j\frac{v_{\eta}^2}{m_N}f(m_{\eta},m_N).
\end{eqnarray}

The contributions from $\Delta \eta=1$ interactions pick up two $\Delta \eta=1$ couplings and two $\eta$ vevs. They obtain negligible contributions becuase of the smallness of $\Delta \eta=1$ couplings. $\Delta\eta =2$ coupling $\lambda_{11}$ can also contribute to $m^{\mbox{break}}$ by picking up $\eta$ vev and it is also significantly small and negligible.

Correcting above all contributions, we find that neutrino mass in our model can be described by the two types of mass structure, $m^{\mbox{sym}}$ and $m^{\mbox{break}}$. 

As we mentioned in Appendix B, under our $A_4$ breaking pattern, the invariance for ($\eta_2,\eta_3$) permutation in scalar potential is crucial to have the special pattern of $m^{\mbox{break}}$ given in Eq.(3.4), that is, to obtain the relation Eq.(3.8). This can be easily seen as follows.  Here notice that our lagragian is invariant for an exchange of ($\eta_2, N_2, (y_{\nu}^{\mu}, L_{\mu}), (y_{\mu}, \mu_R))$ and ($\eta_3, N_3, (y_{\nu}^{\tau}, L_{\tau}), (y_{\tau}, \tau_R))$.~\footnote{If we include $N_5,$, $N_6$, an exchange of ($\eta_2, N_2, (y_{\nu}^{\mu}, L_{\mu}), (y_{\mu}, \mu_R), N_5)$ and ($\eta_3, N_3, (y_{\nu}^{\tau}, L_{\tau}), (y_{\tau}, \tau_R), N_6)$ leaves the lagragian invariant.} In loop diagrams contributing to neutrino masses, we see that fixing the flavors of external two leptons, the amplitudes except for the two vertex with fixed external leptons are invariant against the exchange. The entanglements for the permutation at the two vertex is disentangled by $((y_{\nu}^{\mu}, L_{\mu}), (y_{\nu}^{\eta}, L_{\tau}))$ exchange in the external leptons. As the result, the invariance of scalar potential for $(\eta_2,\eta_3)$ permutation demands the universality for the coefficients of the follwoing two Dim 5 neutrino mass operators,
\begin{eqnarray}
\frac{1}{\Lambda_a}[(y_{\nu}^{\alpha}L)(y_{\nu}^{\beta}L)]_{1'}[(\eta^{\dag} \eta^{\dag})]_{1"},
~\frac{1}{\Lambda_b}[(y_{\nu}^{\alpha}L)(y_{\nu}^{\beta}L)]_{1"}[(\eta^{\dag} \eta^{\dag})]_{1'},
\end{eqnarray}
that is, $\Lambda_a=\Lambda_b$. This universality of the cut-off scale results the relation Eq.(3.8). The relation of Eq.(3.8) is stable against the extentions of models as long as the lagragian is invariant for $(\eta_2,\eta_3)$ permutation.

If the invariance for ($\eta_2,\eta_3$) permutation is lost, e.g by introduing CP phases in scalar potential, since we can not expect a relation such as $\Lambda_a=\Lambda_b$ , the expression for $m^{\mbox{break}}$ is not valid any more and the relation among neutrino mass parameters as Eq(3.8) is lost, which means that we have more freedom to explain neutrino mass.

\end{document}